\documentclass[12pt]{article}
\usepackage{amsmath}
\usepackage{amssymb}
\usepackage{amsthm}
\usepackage{graphicx}
\usepackage{natbib}
\usepackage{url} % not crucial - just used below for the URL
\usepackage{framed}
\usepackage{rotfloat}
\usepackage{multirow}
\usepackage{makecell}
\usepackage[caption=false]{subfig}% Support for small, `sub' figures and tables
\usepackage{epstopdf}% To incorporate .eps illustrations using PDFLaTeX, etc.
\usepackage[usenames]{color}

\DeclareMathOperator*{\argmin}{arg\,min}

\DeclareMathOperator*{\supp}{supp}

\DeclareMathOperator*{\sign}{sign}
\def\prox{{\rm Prox}}
\def\A{{\mathcal A}}

\def\C{{\mathcal C}}

\def\I{{\mathcal I}}

\def\L{{\mathcal L}}

\def\N{{\mathcal N}}
\def\O{{\mathcal O}}

\def\R{{\mathcal R}}
\def\S{{\mathcal S}}

\newtheorem{lemma}{Lemma}[section]

\newtheorem{theorem}{Theorem}[section]
\newtheorem{assumption}{Assumption}[section]
\newtheorem{remark}{Remark}[section]

\newcommand{\blind}{0}

\begin{document}

\def\spacingset#1{\renewcommand{\baselinestretch}%
{#1}\small\normalsize} \spacingset{1}

%%%%%%%%%%%%%%%%%%%%%%%%%%%%%%%%%%%%%%%%%%%%%%%%%%%%%%%%%%%%%%%%%%%%%%%%%%%%%%

\if0\blind
{
  \title{\bf Graph-based Square-Root Estimation for Sparse Linear Regression}
  \author{Peili Li\thanks{
      This work was supported by the National Natural Science Foundation of China under Grant [number 12301420,12471307,12271217,12371289]; the Key Scientific Research Project of Universities in Henan Province [number 25A110002]; the Natural Science Foundation of Henan Province under Grant [number 232300421018]; the National Key R\&D Program of China under Grant [number 2023YFA1008700, 2023YFA1008703]; the Basic Research Project of Shanghai Science and Technology Commission under Grant [number 22JC1400800]; and the Shanghai Pilot Program for Basic Research under Grant [number TQ20220105].}\hspace{.2cm}\\
    School of Mathematics and Statistics,\\ Henan University, Kaifeng 475000, China\\
    %and
    \\
    Zhuomei Li \\
    School of Mathematics and Statistics,\\ Henan University, Kaifeng 475000, China\\
   %     and
   \\
    Yunhai Xiao\thanks{Corresponding author: Yunhai Xiao (yhxiao@henu.edu.cn)} \\
    Center for Applied Mathematics of Henan Province,\\ Henan University, Zhengzhou 450046, China\\
  %      and
  \\
    Chao Ying \\
    Department of Biostatistics and Medical Informatics,\\ University of Wisconsin-Madison, Madison, WI 53726, USA\\
  %      and
  \\
    Zhou Yu \\
    School of Statistics, East China Normal University, Shanghai 200062, China}
  \maketitle
} \fi

\if1\blind
{
  \bigskip
  \bigskip
  \bigskip
  \begin{center}
    {\LARGE\bf Title}
\end{center}
  \medskip
} \fi

\bigskip
\begin{abstract}
Sparse linear regression is one of the classic problems in the field of statistics, which has deep connections and high intersections with optimization, computation, and machine learning. To address the effective handling of high-dimensional data, the diversity of real noise, and the challenges in estimating standard deviation of noise, we propose a novel and general graph-based square-root estimation (GSRE) model for sparse linear regression. Specifically, we use square-root-loss function to encourage the estimators to be independent of the unknown standard deviation of error terms and design a sparse regularization term by using the graphical structure among predictors in a node-by-node form. Based on the predictor graphs with special structure, we highlight the generality by analyzing that the model in this paper is equivalent to several classic regression models. Theoretically, we also analyze the finite sample bounds, asymptotic normality and model selection consistency of GSRE method without relying on standard deviation of error terms. In terms of computation, we employ the fast and efficient alternating direction method of multipliers. Finally, based on a large number of simulated and real data with various types of noise, we demonstrate the performance advantages of the proposed method in estimation, prediction and model selection. Supplementary materials for this article are available online.
\end{abstract}

\noindent%
{\it Keywords:} Sparse linear regression, square-root-loss function, graphical structure among predictors, alternating direction method of multipliers, oracle property.
%\vfill

%\newpage
\spacingset{1.75} % DON'T change the spacing!
\section{Introduction}
\label{sec:intro}

A very active research area in statistics is the linear regression problem, whose classical model is often assumed as $y=X \beta^* +\sigma\epsilon$, where $y\in \mathbb{R}^{n}$ is the observed response variable, $X\in \mathbb{R}^{n\times p}$ is the design matrix, $\beta^{*}\in \mathbb{R}^{p}$ is the true coefficient to be estimated, $\sigma\geq 0$ is the noise level, and $\epsilon\in \mathbb{R}^{n}$ is an error vector of i.i.d. random variables with mean zero. Many practical applications often encounter high-dimensional small-sample data, i.e., situations with small $n$ and large $p$, which will lead to the phenomenon of rank deficiency in the design matrix $X$. Moreover, the effectiveness of traditional least squares estimation model will be greatly challenged when the predictive variables are highly correlated or the data contains non-normal distributed noise. High-dimensional statistical modeling often employs sparse penalty methods, which involves adding different penalty functions to the square loss function $\|y-X \beta\|^2$ for variable selection, such as $\ell_1$ penalty \citep{T1996}, $L_p (0<p<1)$ quasi-norm \citep{FF1993}, smoothly clipped absolute deviation (SCAD) \citep{FL2001}, minimax concave penalty (MCP) \citep{Z2010}, etc. In addition, in some practical problems, such as sparse kernel learning \citep{KY2010}, additive modeling \citep{MVB2009} and multivariate regression \citep{F1991}, variable selection needs to be performed in the form of groups, which has led to the development of group penalty methods, such as group lasso \citep{BV2011,YL2006}. However, traditional group penalty methods often require non-overlapping structure between subgroups, which makes them unsuitable for the practical problems where features may be encoded in multiple groups, such as many biological studies on gene interactions \citep{LL2008,STM2005}.

To effectively handle group sparsity problems with overlapping structures between subgroups, we have noticed that some literature has utilized the connectivity of an undirected graph to simulate the structural information among the predictors and improve sparse regression models based on this graph. Early literature mostly used edge-by-edge method, which employs penalty functions to promote similarity in regression coefficients for correlated predictors \citep{BR2008,KX2009,LL2008,YYLSWY2012}. However, these methods did not directly utilize the neighborhood information of the graph. Additionally, when there are more edges in the graph, this approach leads to more complex penalty terms. As an improvement, \citet{YL2016} proposed a node-wise penalty by using a node-by-node method. Specifically, given the predictor graph $G$ and positive weights $\tau_i$'s, they defined the following norm
\begin{align}\label{gnorm}
\big\|\beta\big\|_{G,\tau}=\min\limits_{\sum_{i=1}^{p}V^{(i)}=\beta, \supp(V^{(i)})\subseteq \mathcal{N}_{i}}\sum_{i=1}^{p}\tau_i \|V^{(i)}\|_2,
\end{align}
where $V^{(i)}\in \mathbb{R}^{p}$ is a partition of $\beta$, and there may be overlapping parts among the $V^{(i)}$'s, $\supp(V^{(i)})$ is the support of vector $V^{(i)}\in \mathbb{R}^{p}$, $\|\cdot\|_2$ is the $\ell_2$ norm, and $\mathcal{N}_{i}$ represents the set containing the predictor $i$ and its neighbors in graph $G$. To avoid repetition, the design motivation of $\|\beta\|_{G,\tau}$ is summarized in the supplementary materials. They no longer focus on the original variable $\beta$ of the model, but decompose $\beta$ into $p$ parts, and then use group sparsity idea based on predictor graph information. With the above notation, their optimization model in sparse regression incorporating
graphical structure among predictors (SRIG) is given by
\begin{equation}\label{YLmodel}
\min_{\beta\in\mathbb{R}^p}\Big\{\frac{\|y-X \beta\|_{2}^2}{2 n}+\lambda \|\beta\|_{G,\tau}\Big\},
\end{equation}
where $\lambda$ is a regularization parameter. The SRIG method facilitates the inclusion of other predictors connected to an important predictor in a model, which is better suited to meet the demands of practical applications. Compared to penalty methods that do not use the graphical structure among predictors, SRIG method can significantly improve the performance of variable selection, estimation and prediction. The good performance of this research idea has attracted the attention of many scholars and has been extended to linear discriminant analysis \citep{LYL2019}, generalized linear regression \citep{ZZZ2019,LL2022,SAD2021}, expectile regression \citep{PZW2023}, quantile regression \citep{WLT2021}, etc.

Despite these attractive features, there still exists a theoretical and practical challenge. Because the aforementioned problems based on the smooth square loss usually need to assume that the noise follows a normal distribution, and the estimation performance depends on knowing standard deviation of the error terms in advance. The theoretical form of the tuning parameter $\lambda$, which is crucial for the model estimation, relies on the standard deviation $\sigma$. However, in high-dimensional settings, accurately estimating $\sigma$ is a non-trivial task, and the types of noise generally exhibit diversity. Fortunately, the proposal of the square-root-loss function overcomes the aforementioned limitations, which was first proposed by \citet{O2007}. Based on the idea of square-root loss, \citet{BCW2011} extended the traditional lasso problem and obtained the following square-root lasso (SRL):
\begin{equation}\label{srlmodel1}
\min_{\beta\in\mathbb{R}^p}\Big\{\frac{\|y-X \beta\|_{2}}{\sqrt{n}}+\frac{\lambda}{n} \|\beta\|_1\Big\},
\end{equation}
where $\|\beta\|_1=\sum_{i=1}^{p}|\beta_i|$ is the $\ell_1$ norm of $\beta$. It has been theoretically validated that SRL can achieve near-oracle property without knowing or estimating $\sigma$ in advance. This makes the lasso-type methods much more attractive in high-dimensional scenarios. Subsequently, the square-root loss was shown to be applicable to non-normal noise under reasonable conditions \citep{BLT2018,CPS2018}. Furthermore, \citet{BLS2013} extended the ideas behind the SRL for group selection. \citet{TWST2020} proposed a nonconvex square-root-loss regression model. For more research on the square-root-loss function, one can refer to \citet{D2018}, \citet{DZLX2021} and \citet{SCG2016}.

In this paper, we carry out research on integrating the square-root-loss regression method and the penalty methods based on graphical structure among predictors. This research aims to construct a novel model for the classic high-dimensional linear regression problem, which can analyze the intrinsic information in data. Specifically, we propose the following graph-based square-root estimation (GSRE) model:
\begin{equation}\label{pmodel1}
\min_{\beta\in\mathbb{R}^p}\Big\{\frac{\|y-X \beta\|_{2}}{\sqrt{n}}+\frac{\lambda}{n} \|\beta\|_{G,\tau}\Big\}.
\end{equation}
Firstly, for several classes of predictor graphs with special structures, we verify the equivalence of the proposed GSRE method with several existing methods. Then, when the graphical structure $G$ of the predictors is known, we provide finite sample bounds, asymptotic normality and model selection consistency for the GSRE method with a $\sigma$-free tuning sequence $\lambda$. For the case where the graphical structure $G$ is unknown, we construct it by estimating sparse precision matrix of the predictors \citep{CLL2011,FHT2008,YL2007}. In terms of computation, we adopt the classical alternating direction method of multiplies (ADMM) in the optimization field. Simulation studies and practical applications demonstrate that the proposed GSRE method exhibits good performance in variable selection, model estimation and prediction.

To the best of our knowledge, this is the first study to introduce the graphical structure among predictors into square-root-loss regression. The theoretical analysis relies on a $\sigma$-free tuning parameter, which completely eliminates the dependence on the standard deviation of the noise. Additionally, the results of this paper are also applicable to scenarios with non-Gaussian noises, which is more in line with practical needs. Furthermore, for predictors with complex correlation structures, the ingenious design of penalty terms allows for in-depth analysis and effective processing of the essential correlations. These innovations jointly expand the scope of practical applications of the research results, and provide more generalizable technical support for regression analysis in complex data environments.

\textbf{Organization:} The remainder of this paper is organized as follows. Relevant and rigorous theoretical results are presented in Section \ref{the}. Section \ref{alg} provides a detailed description of the classical ADMM for model (\ref{pmodel1}). Section \ref{num} demonstrates the numerical performance of the proposed GSRE method on both simulated and real data. Finally, we summarize this paper in Section \ref{con}. Technical proofs are provided in the supplementary materials.

\textbf{Notation:} To end this section, we summarize some notations used in this paper. $[p]:=\{1,2,\ldots,p\}$. When $|\cdot|$ is applied to a scalar, it denotes the absolute value, and when it acts on a set, it denotes the cardinality. For each $i \in [p]$, we denote $d_i$ as the number of predictors in the neighborhood $\N_i$, that is, $d_i=|\N_i|$. $d_{min}:=\min_{1\leq i\leq p}d_i$, $d_{max}:=\max_{1\leq i\leq p}d_i$, $\tau_{min}:=\min_{1\leq i\leq p}\tau_i$. For each $i\in [p]$, denote $\zeta_i=\|X_{\N_i}\|^2$ and $\zeta=\max_{i}\zeta_i$, where $\|X_{\N_i}\|$ is the operator norm of the matrix $X_{\N_i}$. For a $n\times p$ matrix $M$, $\|M\|_\infty$ is defined as $\max_{1\leq i\le n}\sum_{j=1}^{p}|M_{ij}|$.
%{\color{red}For any random variable $A$, we define the norm $\|A\|_{\Psi_1}:=\inf\{t>0 : E (\exp(\frac{|A|}{t}))\leq 2\}$.}
For a constant $x$, we denote $(x)_+=\max(x,0)$. Here, we denote $\I_{*}:=\{i: \beta_i^{*}\neq 0\}$, $\I_{*}^{c}:=\{i: \beta_i^{*}= 0\}$ and $s^*=|\I_{*}|$ as the true nonzero coefficient set, the true zero coefficient set, and the number of true nonzero coefficients, respectively. For every $\beta\in\R^{p}$, denote $\beta_{\I_*}$ and $\beta_{\I_*^c}$ as the subvectors of $\beta$ with indices in $\I_*$ and $\I_*^c$ respectively. For any matrix $M$, $M_{\I_*,\I_*}$ denote the submatrix of $M$ consisting of the entries with row and column indices in $\I_*$. $\beta_{min}^ *=\min_{i\in \I_*}|\beta_i^*|$ denote the minimum absolute nonzero coefficient. $\sign(\cdot)$ maps a positive entry to $1$, a negative entry to $-1$, and zero to zero. The symbol $\delta_{\C}(x)$ represents the  indicator function over $\C$ such that $\delta_{\C}(x)=0$ if $x\in\C$ and $+\infty$ otherwise.

\section{Theoretical results}\label{the}
%\setcounter{equation}{0}
%%%%%%%%%%%%%%%%%%%%%%%%%%%%%%%%%%%%%%%%%%%%%%%%%%%%%%
In this section, we will give some relevant and rigorous theoretical results with a non-random design matrix. Specifically, we first characterize the features of optimal solution by analyzing the subgradient conditions of GSRE problem. Then, based on the given predictor graphs with certain special structures, we establish the connections between the proposed GSRE model and several other existing penalized problems. Subsequently, we analyze the finite sample bounds, asymptotic normality and model selection consistency of the GSRE method with a $\sigma$-free tuning sequence $\lambda$.
\subsection{Subgradient conditions}
\begin{lemma}\label{sc}
 Assume there exists a vector $\beta\in \mathbb{R}^{p}$ satisfying $X \beta\neq y$. Then $\beta$ is an optimal solution of (\ref{pmodel1}) if and only if $\beta$ can be decomposed as $\beta=\sum_{i=1}^{p}V^{(i)}$ where $V^{(i)}$'s satisfy that, for all $1\leq i \leq p$,
\begin{enumerate}
  \item[(a)] $V_{\N_{i}^{c}}^{(i)}=0$;
  \item[(b)] either $V_{\N_{i}}^{(i)}\neq 0$ and $\frac{X_{\N_i}^{\top}(y-X \beta)}{\|y-X \beta\|_2}=\frac{\lambda \tau_i V_{\N_{i}}^{(i)}}{\sqrt{n}\|V_{\N_{i}}^{(i)}\|_2}$, or $V_{\N_{i}}^{(i)}=0$ and $\frac{\|X_{\N_i}^{\top}(y-X \beta)\|_2}{\|y-X \beta\|_2}\leq \frac{\lambda \tau_i}{\sqrt{n}}$.
\end{enumerate}
\end{lemma}
%\begin{proof}
%See Appendix \ref{sc1}.
%\end{proof}
\begin{remark}
According to Lemma \ref{sc}, if $(\hat{V}^{(1)},\hat{V}^{(2)},\ldots,\hat{V}^{(p)})$ is a decomposition of the optimal solution of (\ref{pmodel1}), then for each $i$, either $\hat{V}^{(i)}=\textbf{0}_{p}$ or $\supp(\hat{V}^{(i)})=\N_i$. This result is very consistent with the design idea of node-wise penalty.
\end{remark}
\subsection{Connections with some existing methods}
\begin{lemma}\label{lemma2}
\begin{enumerate}
  \item[(a)] If the predictor graph has no edge, the problem (\ref{pmodel1}) is the same as
  \begin{equation}\label{al}
   \min_{\beta\in\mathbb{R}^p}\Big\{\frac{\|y-X \beta\|_{2}}{\sqrt{n}}+\frac{\lambda}{n} \sum_{i=1}^{p}\tau_i |\beta_i|\Big\}.
  \end{equation}
  When $\tau_i\equiv1$ for each $i$, (\ref{al}) is the classic square-root lasso \citep{BCW2011}.
  \item[(b)] If the predictor graph consists of $J$ disconnected complete subgraphs, and the nodes in the $J$ disconnected complete subgraphs are assumed to be $\{1,2,\ldots,q_1\}, \{q_1+1,\ldots,q_1+q_2\},\ldots, \{q_{J-1}+1,\ldots,q_{J-1}+q_J\}$, then the problem (\ref{pmodel1}) is equivalent to
  \begin{equation}\label{gl}
   \min_{\beta\in\mathbb{R}^p}\Big\{\frac{\|y-X \beta\|_{2}}{\sqrt{n}}+\frac{\lambda}{n} \sum_{j=1}^{J}\tilde{\tau}_j \|\beta^{(j)}\|_2\Big\},
  \end{equation}
  where $\tilde{\tau}_j=\min\limits_{q_{j-1}+1\leq i \leq q_{j-1}+q_j}\tau_i$ and $\beta^{(j)}=(\beta_{q_{j-1}+1}, \beta_{q_{j-1}+2}, \ldots, \beta_{q_{j-1}+q_j})^{\top}$. Obviously, this problem is exactly the group square-root lasso problem \citep{BLS2013}.
  \item[(c)] If the predictor graph is a complete graph, then the problem (\ref{pmodel1}) has the same nonzero solutions as
  \begin{equation}\label{rl}
   \min_{\beta\in\mathbb{R}^p}\Big\{\frac{\|y-X \beta\|_{2}}{\sqrt{n}}+\frac{\tilde{\lambda}}{n}\|\beta\|_2^2\Big\},
  \end{equation}
  where $\tilde{\lambda}$ is a tuning parameter different from $\lambda$.
\end{enumerate}
\end{lemma}
%\begin{proof}
%See Appendix \ref{cwe1}.
%\end{proof}
\begin{remark}
Lemma \ref{lemma2} shows that square-root lasso and group square-root lasso are special cases of the proposed model (\ref{pmodel1}). This conclusion validates the broader applicability of our model, enabling it to handle predictor graphs with diverse structures.
\end{remark}
\subsection{Choice of the tuning parameter}

As is well known, accurately estimating the unknown standard deviation $\sigma$ of noise is very challenging in high-dimensional settings. In this part, we will use the ratio of random variables to determine the choice of the tuning parameter $\lambda$, thus removing the dependence on $\sigma$. Specifically, for any predictor graph G, we define
$$V:=\max\limits_{1\leq i\leq p}\Big\{ \frac{\sqrt{n}\|X_{\N_i}^{\top}\epsilon\|_2}{\tau_i\|\epsilon\|_2}\Big\}.$$
For a given constant $r>1$, we set $\bar{r}:=\frac{r+1}{r-1}$. Given $\lambda>0$, we define
\begin{align}\label{setA}
	\A:=\bigl\{ V\le\lambda/\bar{r}  \bigr\},
\end{align}
which represents a event that inequality $V\le\lambda/\bar{r}$ holds. Next, we will prove that, for any $\alpha$ close to zero and a suitable tuning parameter $\lambda$, the event $\A$ holds with probability $1-\alpha$. Before presenting the specific conclusion, we first provide the following assumption:
\begin{assumption}\label{ass1}
The additive noise terms $\epsilon_i$ follow a standard Gaussian distribution.
\end{assumption}
%%%%%%%%%%%%%%%%%%%%%%%%%%%%%%%%%%%%%%%%%%%%%%%%%%%%%%%%%%%%%%%%%%%%%%%%%%%%%%%%%%%%%%%%%%%%%%
%%%%%%%%%%%%%%%%%%%%%%%%%%%%%%%%%%%%%%%%%%%%%%%%%%%%%%%%%%%%%%%%%%%%%%%%%%%%%%%%%%%%%%%%%
\begin{lemma}\label{Vle}
	Assumption \ref{ass1} is satisfied and $0<d_i<n$ for all $1\leq i\leq p$. Let $\alpha\in (0,1)$ be given such that $16 \log(2p/\alpha)\leq n-d_{max}$. Then if we choose a parameter $\lambda$ that satisfies
	\begin{align}\label{lam}
	\lambda\ge\frac{n\bar{r}\sqrt{2\zeta d_{max}}}{\tau_{min}\sqrt{n-d_{max}}}\Big(1+\sqrt{\frac{2log(2p/\alpha)}{d_{min}}}\Big),
	\end{align}
	it holds that
	$\mathbb{P}(V\leq \frac{\lambda}{\bar{r}})\geq 1-\alpha$.
\end{lemma}
%\begin{proof}
%See Appendix \ref{lamp}.
%\end{proof}

\begin{remark}
Based on $d_i=|\N_i|$, for any $i$, we can have $d_i\geq 1$. Additionally, since we are concerned with sparsity issues, it is a common assumption that the number of mutually related predictors in the predictor graph is less than $n$. An immediate consequence of this lemma is that, for any given $\alpha$ approaching 0, a $\sigma$-free expression of $\lambda$ holds with probability $1-\alpha$, which is a very important additional benefit.
\end{remark}
%%%%%%%%%%%%%%%%%%%%%%%%%%%%%%%%%%%%%%%%%%%%%%%%%%%%%%%%%%%%%%%%%%%%%%%%%%%%%%%%%%%

\begin{remark}
It can be noted that the analysis in this part only considers Gaussian noise, but the results can also be generalized to different types of noise by applying different deviation inequalities \citep{B2002,LV2011,VL2013}. For example, if $\epsilon_i$'s belong to a general sub-exponential family, the order of magnitude of $\lambda$ remains the same. The corresponding analysis can be found in the supplementary materials. For more analysis on non-Gaussian noise, one can refer to \citet{BCW2011}. A detailed study is omitted here.
\end{remark}

\subsection{Finite sample bounds}

In this subsection, we will give the results of finite sample bounds for the proposed GSRE method over the event $\A$. The following assumptions need to be considered in subsequent analysis.
\begin{assumption}\label{ass2}
The neighborhood $\N_i\subseteq \I_*$ for each $i\in \I_*$.
\end{assumption}
\begin{assumption}\label{ass3}
For a subset $J\subset [p]$, there exists $\kappa >0$ and $r>1$ such that
      \begin{align*}
      \sum_{i\in J}\tau_i\|V^{(i)}\|_2\leq \frac{\sqrt{s^*}\|X\beta\|_2}{\sqrt{n}\kappa}
      \end{align*}
  for all $(V^{(1)},V^{(2)},\ldots,V^{(p)}) \in \mathcal{T}_{G,\tau}(\beta,J)$, where $\mathcal{T}_{G,\tau}(\beta,J)$ is the set of all optimal decompositions of $\beta$ such that $$\sum_{i\in J^c}\tau_i\|V^{(i)}\|_2\leq r \sum_{i\in J}\tau_i\|V^{(i)}\|_2.$$
\end{assumption}
\begin{assumption}\label{ass4}
The number of true nonzero coefficients $s^*$ satisfies $s^*<\frac{n^2 \kappa^2}{\lambda^2}$.
\end{assumption}
\begin{remark}
Assumption \ref{ass2} guarantees that predictors connected to the useful predictors are also useful, which is reasonable. Assumption \ref{ass3} is a general compatibility condition on the design matrix \citep{BV2011,BLS2013} and is a slight relaxation of the widely used restricted eigenvalue condition \citep{BRT2009}. The constant $\kappa$ quantifies the correlations in the design matrix: a smaller value of $\kappa$ indicates stronger correlations. An example where Assumption \ref{ass3} holds is presented in the supplementary materials. Assumption \ref{ass4} essentially constrains the true model to have fewer parameters than $n$.
\end{remark}

%%In addition, for each $j\in \{1,2,\cdots,p\}$, denote $\zeta_j=\|X_{\N_j}\|^2$ and $\zeta=\max\limits_{j}\zeta_j$, where $\|X_{\N_j}\|$ is the operator norm of the matrix $X_{\N_j}$.
%%%%%%%%%%%%%%%%%%%%%%%%%%%%%%%%%%%%%%%%%%%%%%%%%%%%%%%%%%%%%%%%%%%%%%%%%%%%%%%%%%%%%%%%
\begin{theorem}\label{th1}
 Assumptions \ref{ass1}, \ref{ass2}, \ref{ass3}, \ref{ass4} hold. Then, on the event $\A$, for any optimal solution $\hat{\beta}$ of model (\ref{pmodel1}), we have
\begin{align*}
\frac{\|X(\hat{\beta}-\beta^*)\|_2}{\sqrt{n}}\lesssim  \frac{\sigma \lambda \sqrt{s^*}}{n\kappa},\quad
\|\hat{\beta}-\beta^*\|_{G,\tau}\lesssim \frac{\sigma \lambda s^*}{n\kappa^2},\quad
\|\hat{\beta}-\beta^*\|_2\lesssim \frac{\sigma \lambda s^*}{n\kappa^2\tau_{min}},
\end{align*}
with probability at least $1-(1+e^2)e^{-n/24}$.
\end{theorem}
%\begin{proof}
%See Appendix \ref{th11}.
%\end{proof}

%%%%%%%%%%%%%%%%%%%%%%%%%%%%%%%%%%%%%%%%%%%%%%%%%%%%%%%%%%%%%%%%%%%%%%%%%%%%%%%%%%%%
%%%%%%%%%%%%%%%%%%%%%%%%%%%%%%%%%%%%%%%%%%%%%%%%%%%%%%%%%%%%%%%%%%%%%%%%
\begin{remark}
Theorem \ref{th1} gives similar results to group lasso and group square-root lasso estimators \citep{LPV2011, BLS2013}. It is worth noting that the estimation analysis is based on a $\sigma$-free tuning parameter, which is obviously different from SRIG in \citet{YL2016}. Furthermore, it is noted that, from Theorem \ref{th1}, subset recovery can be guaranteed with no noise, i.e., $\sigma=0$. However, for classical least squares methods such as lasso, elastic net and group lasso, even if $\sigma=0$, accurate recovery cannot be guaranteed because their tuning parameters are heavily dependent on $\sigma$. In addition, an empirical verification of $\|\hat{\beta}-\beta^*\|_2$ can be found in the supplementary materials. Additionally, combining with Lemma \ref{Vle} on the selection of regularization parameter, it can be concluded that the result of Theorem \ref{th1} holds with probability $1-\alpha-(1-\alpha)(1+e^2)e^{-n/24}$.
%\begin{corollary}
%	Assume that condition (A1) is satisfied and $n-\tau_{max}^2>0$.Let $\alpha\in (0,1)$ be given such that $16 \log(\frac{2p}{\alpha})\leq n-\tau_{max}^2$.
%	\item (i) Under the conditions of theorem \ref{th1}, its conclusion holds with probability at least $1-\alpha-(1-\alpha)(1+e^2)e^{-n/24}$.
%	\item (ii) Under the conditions of theorem \ref{th3}, its conclusion holds with probability at least $1-\alpha$.
%\end{corollary}
\end{remark}
\subsection{Asymptotic normality and model selection consistency}
In this subsection, we first study the asymptotic normality of the proposed GSRE method in the case where the dimension $p$ is fixed. Then, we analyze the model selection consistency in the high-dimensional case where $p$ grows with $n$. For the case of fixed $p$, we need the following two commonly used assumptions:

\begin{assumption}\label{ass5}
As $n \to \infty $, $X^\top X/n \to M$, where $M$ is a positive matrix.
\end{assumption}
\begin{assumption}\label{ass6}
The errors  $\epsilon_1$, $\epsilon_2$, $\ldots$, $\epsilon_n$ are i.i.d. variables with mean 0 and variance $1$.
\end{assumption}
%%%%%%%%%%%%%%%%%%%%%%%%%%%%%%%%%%%%%%%%%%%%%%%%%%%%%%%%%%%%%%%%%%%%%%%%%%%%%%%%%%%%%%%%%
\begin{theorem}\label{th2}
Assumptions \ref{ass2}, \ref{ass5}, \ref{ass6} hold. Suppose that as $n\rightarrow \infty$, the tuning parameter $\lambda$ is chosen such that $\lambda/\sqrt{n}\to 0$ and $\lambda n^{(\nu-1)/2}\to\infty$ for some $\nu>1$. Furthermore, $\tau_i=O(1)$ for each $i \in \I_*$ and $\lim\inf_{n\to\infty}n^{-\nu/2}\tau_i>0$ for each $i \in \I_*^c$. Then, with dimension p fixed, as $n\to\infty$, we have
	\begin{align}\label{th2g1}
	\sqrt{n}(\hat{\beta}_{\I_*}-\beta_{\I_*}^*)\overset{d}{\rightarrow} N(0,\sigma^2{M}^{-1}_{\I_*,\I_*}),\quad {\hat\beta}_{\I_*^c}\overset{p}{\rightarrow}0.
	\end{align}
\end{theorem}
%\begin{proof}
%See Appendix \ref{th21}.
%\end{proof}
%%%%%%%%%%%%%%%%%%%%%%%%%%%%%%%%%%%%%%%%%%%%%%%%%%%%%%%%%%%%%%%%%%%%%%%%%%%%%%%%%%%%%%%%%
\begin{remark}
Theorem \ref{th2} indicates that the proposed GSRE method is estimation-consistent when $p$ is fixed. Clearly, the above result can be naturally extended to the random design setting. The relevant detailed discussion is omitted here.
\end{remark}

\begin{remark}
Theorem \ref{th2} also provides a guideline on how to choose the positive weight $\tau_i$. \citet{OJV2011} suggest to consider $\tau_i=d_i^\eta$ for some $\eta\in\big(0,\frac{1}{2}\big)$. More specially, they give a critical value with $\eta=\frac{log(2)}{2log(3)}$, which is the smallest value that it becomes impossible to select a support of exactly two covariates. In our numerical experiments, we also choose $\tau_i=d_i^\eta$ with $\eta=\frac{log(2)}{2log(3)}$.
\end{remark}

Next, we consider the high-dimensional case where the dimension $p$ grows with $n$, while the design matrix $X$ is fixed. The subsequent analysis requires the following assumptions:
\begin{assumption}\label{ass7}
The number of nonzero coefficients $s^*=O(n^{\delta_0})$ for some constant $\delta_0 \in (0,1)$.
\end{assumption}
\begin{assumption}\label{ass8}
There exists a constant $Q_1>0$ such that $\max_{i\in \I^c_*}\|X_i\|_2\le \frac{Q_1}{\sqrt{n}}$ for each $n$.
\end{assumption}
\begin{assumption}\label{ass9}
There exists a constant $Q_2>0$ such that the smallest eigenvalue of $\frac{X_{\I_*}^\top X_{\I_*}}{n}$ is larger than $Q_2$ for each $n$.
\end{assumption}
\begin{assumption}\label{ass10}
There exists a constant $\xi \in (0,1)$ such that $\|X^{\top}_{\N_i}X_{\I_*}(X^{\top}_{\I_*}X_{\I_*})^{-1}\|_\infty\le1-\xi$.
\end{assumption}
%%%%%%%%%%%%%%%%%%%%%%%%%%%%%%%%%%%%%%%%%%%%%%%%%%%%%%%%%%%%%%%%
\begin{remark}
Assumption \ref{ass7} ensures the sparsity of the problem. Assumption \ref{ass8} can be achieved by normalizing each predictor. Assumption \ref{ass9} guarantees that matrix $\frac{X_{\I_*}^\top X_{\I_*}}{n}$ is invertible. Assumption \ref{ass10} is similar to the strong irrepresentable condition \citep{ZY2006}.
\end{remark}
\begin{theorem}\label{th3}
	Assumptions \ref{ass1}-\ref{ass4} and \ref{ass7}-\ref{ass10} hold. Suppose $\tau_i=\sqrt{d_i}m_i$ for each $i$, where the $m_i$'s satisfy that $\max_{i\in \I_*}m_i=O_p(1)$ and $lim \inf_{n\to\infty}n^{-\varsigma}\min_{i\in \I_*^c}m_i>0$ for some $\varsigma > \delta_0$. Furthermore, as $n\to\infty$ and $p=p(n)\to\infty$, the selected tuning parameter $\lambda$ satisfy that,
$$\frac{1}{\beta_{min}^*}\Big(3\sigma\sqrt{\frac{log s^*}{nQ_2}}+\frac{2\sigma\lambda\sqrt{s^*}}{nQ_2}\big(1+\frac{C\lambda \sqrt{s^*} }{n\kappa}\big)\max_{i\in \I_*}\tau_i\Big)\to 0 \ \text{and} \ \frac{\sqrt{log(p-s^*)}}{\lambda\sqrt{n}\min\limits_{i\in \I_*^c}m_i}\to 0.$$
Then, on the event $\A$, as $n\to \infty$ and $p=p(n)\to\infty$, there exists a solution $\hat{\beta}$ to (\ref{pmodel1}) such that $\sign(\hat{\beta})=\sign(\beta^*)$ with probability tending to 1.
\end{theorem}
%\begin{proof}
%See Appendix \ref{th31}.
%\end{proof}
\begin{remark}
	Theorem $\ref{th3}$ indicates that the proposed GSRE method is model selection consistent in high-dimensional settings.
\end{remark}

\section{Computation}\label{alg}
%\setcounter{equation}{0}
%%%%%%%%%%%%%%%%%%%%%%%%%%%%%%%%%%%%%%%%%%%%%%%%%%%%%%

Upon careful observation, we note that model (\ref{pmodel1}) has two non-smooth convex structures. It is well known that ADMM is a flexible and efficient optimization method that can handle convex optimization problems with separable structures and linear constraints. Therefore, we use ADMM for the computation here. First, to effectively utilize ADMM, we introduce two auxiliary variables to reformulate model (\ref{pmodel1}) as the following minimization problem with two-block separable structures and two linear constraints:
\begin{equation}\label{pmodel2}
\begin{array}{lll}
\min \limits_{\beta, u, v} & \frac{\|u\|_{2}}{\sqrt{n}}+\frac{\lambda}{n} \|v\|_{G,\tau} \\
\text{s.t.} & X \beta-u=y, \\
& \beta-v=0.
\end{array}
\end{equation}
Given $\sigma>0$, the augmented Lagrangian function of (\ref{pmodel2}) is defined by
\begin{align*}
\L_{\sigma}(\beta,u,v;z,w):=&\frac{\|u\|_{2}}{\sqrt{n}}+\frac{\lambda}{n}\|v\|_{G,\tau}+\langle z,X \beta-u-y\rangle +\langle w,\beta-v\rangle\\
&+\frac{\sigma}{2}\|X \beta-u-y\|^2+\frac{\sigma}{2}\|\beta-v\|^2.
\end{align*}
Starting from an initial point $(u^0,v^0,z^0,w^0)\in \mathbb{R}^n \times \mathbb{R}^p\times \mathbb{R}^n \times \mathbb{R}^p$, we use ADMM to update the variables sequentially. Specifically, with the given $(u^k,v^k,z^k,w^k)$, the new iteration $(\beta^{k+1},u^{k+1},v^{k+1},z^{k+1},w^{k+1})$ is generated via the following iterative scheme:
\begin{equation}\notag
\left\{
\begin{array}{llll}
\beta^{k+1}=\argmin\limits_{\beta\in \mathbb{R}^p} \L_{\sigma}(\beta,u^k,v^k;z^k,w^k),\\[3mm]
u^{k+1}=\argmin\limits_{u\in \mathbb{R}^n} \L_{\sigma}(\beta^{k+1},u,v^{k};z^k,w^k),\\[3mm]
v^{k+1}=\argmin\limits_{v\in \mathbb{R}^p} \L_{\sigma}(\beta^{k+1},u^{k+1},v;z^k,w^k) ,\\[3mm]
z^{k+1}=z^k+\tau \sigma (X \beta^{k+1}-u^{k+1}-y),\\[3mm]
w^{k+1}=w^k+\tau \sigma(\beta^{k+1}-v^{k+1}),
\end{array}
\right.
\end{equation}
where $\tau\in \big(0,(1+\sqrt{5})/2\big)$ is the step size. Next, we will primarily focus on solving the inner minimization subproblems.

First, for the $\beta$-subproblem, we can easily get
\begin{align*}
\beta^{k+1}&=\argmin\limits_{\beta\in \mathbb{R}^p} \Big\{\langle z^k,X \beta-u^k-y\rangle +\langle w^k,\beta-v^k\rangle+\frac{\sigma}{2}\|X \beta-u^k-y\|^2+\frac{\sigma}{2}\|\beta-v^k\|^2\Big\}\\
&=\Big(\I_{p}+X^{\top}X\Big)^{-1}\Big(v^k-\sigma^{-1}w^k+X^{\top}(u^k+y-\sigma^{-1}z^k)\Big).
\end{align*}
In order to avoid the time-consuming issue of inverting large matrices in high-dimensional environments, we use the Sherman-Morrison-Woodbury formula \citep{GV1996} in specific computations, i.e., $\Big(\I_{p}+X^{\top}X\Big)^{-1}=\I_p-X^{\top}(\I_n+XX^{\top})^{-1}X$. This process successfully transfers the inverse operation of a $p\times p$ matrix to a $n\times n$ matrix, which greatly reduces the computational cost of the algorithm when $p$ is much larger than $n$.

For the $u$ and $v$ subproblems, with the help of proximal point mapping, we can obtain
\begin{align*}
&u^{k+1}=\argmin\limits_{u\in \mathbb{R}^n} \Big\{\frac{\|u\|_{2}}{\sqrt{n}}-\langle z^k,u\rangle+\frac{\sigma}{2}\|X \beta^{k+1}-u-y\|^2 \Big\}=\prox_{\frac{1}{\sigma \sqrt{n}}\|\cdot\|_{2}}(X\beta^{k+1}-y+\sigma^{-1}z^k).\\
&v^{k+1}=\argmin\limits_{v\in \mathbb{R}^p} \Big\{\frac{\lambda}{n}\|v\|_{G,\tau}-\langle w^k,v\rangle+\frac{\sigma}{2}\|\beta^{k+1}-v\|^2\Big\}=\prox_{\frac{\lambda}{\sigma n} \|\cdot\|_{G,\tau}}(\beta^{k+1}+\sigma^{-1}w^k).
\end{align*}
Denote $r^{k+1}:=X\beta^{k+1}-y+\sigma^{-1}z^k$. From \citet[Example 6.19]{B2017}, we have
$$\prox_{\frac{1}{\sigma \sqrt{n}}\|\cdot\|_{2}}(r^{k+1})=\Big(1-\frac{1}{max\{\sigma \sqrt{n}\|r^{k+1}\|_2,1\}}\Big)r^{k+1}.$$
Let $h^{k+1}:=\beta^{k+1}+\sigma^{-1}w^{k}$. To compute the proximal point of $h^{k+1}$ associated with $\frac{\lambda}{\sigma n} \|\cdot\|_{G,\tau}$, we first give the following result:
\begin{lemma}\label{progt}
	If $f(v)=\frac{\lambda}{\sigma n}\|v\|_{G,\tau}$, we have
	$$\prox_{f}(x)=x-\Pi_{\S_{\mathcal{O}}}(x),$$
	where $\mathcal{O}:=\{i\in [p]\ | \ \|x_{\N_i}\|_2>\frac{\lambda\tau_i}{\sigma n}\}$, $\S_{\mathcal{O}}:=\{x\in \mathbb{R}^p \ | \ \|x_{\N_i}\|_2\leq \frac{\lambda\tau_i}{\sigma n} \text{ for each } i\in \mathcal{O}\}$.	
\end{lemma}
%\begin{proof}
%	See Appendix \ref{progt1}.
%\end{proof}

Then, it is easy to see that
$$\prox_{\frac{\lambda}{\sigma n}\|\cdot\|_{G,\tau}}(h^{k+1})=h^{k+1}-\Pi_{S_{\O^{k+1}}}(h^{k+1}),$$
where $\O^{k+1}:=\left \{i\in [p] \ | \ \|h^{k+1}_{\N_i}\|_2>\frac{\lambda\tau_i}{\sigma n} \right\}$, $S_{\O^{k+1}}:=\left \{x\in\Bbb R^p \ | \ \|x_{\N_i}\|_2\le\frac{\lambda\tau_i}{\sigma n} \text{ for each } i\in \O^{k+1}\right \}$. The main problem we face is to solve the projection of $h^{k+1}$ onto the convex set $S_{\mathcal{O}^{k+1}}$. Similar to \citet{YL2016}, we also use different methods to calculate flexibly based on the number of elements in $\O^{k+1}$, denoted as $|\O^{k+1}|$. Specifically, when $|\O^{k+1}|< p/10$, we use the Bertseka's projected Newton method \citep{SL2014} to calculate this projection from the dual perspective. However, when $|\O^{k+1}|\geq p/10$, we use the Parallel Dykstra-like proximal algorithm \citep{CP2011}. The details about these two algorithms are
shown in the supplementary materials of \citet{YL2016}.

Based on the above analysis, we see that each minimization subproblem has been effectively computed, and thus we can summarize the algorithm framework of ADMM for model (\ref{pmodel2}).
%%%%%%%%%%%%%%%%%%%%%%%%%%%%%%%%%%%%%%%%%%%%%%%%%%%%%%%%%%
\begin{framed}
	\noindent
	{\bf ADMM Algorithm}
	\vskip 2.0mm \hrule \vskip 4mm
	\noindent
	{\bf Step 0}: Let $\sigma>0$, $\tau\in(0,(1+\sqrt{5})/2)$ be given parameters. Choose $(u^0,v^0,z^0,w^0)\in \mathbb{R}^n \times \mathbb{R}^p\times \mathbb{R}^n \times \mathbb{R}^p$. For $k=0,1,\ldots$, perform the following steps iteratively:\\
	{\bf Step 1}: Given $u^k$, $v^k$, $z^k$ and $w^k$, compute $$\beta^{k+1}=\Big(\I_{p}+X^{\top}X\Big)^{-1}\Big(v^k-\sigma^{-1}w^k+X^{\top}(u^k+y-\sigma^{-1}z^k)\Big).$$
	{\bf Step 2}: Given $\beta^{k+1}$, $z^k$ and $w^k$, compute $$u^{k+1}=\prox_{\frac{1}{\sigma \sqrt{n}}\|\cdot\|_{2}}(X\beta^{k+1}-y+\sigma^{-1}z^k), \quad v^{k+1}=\prox_{\frac{\lambda}{\sigma n} \|\cdot\|_{G,\tau}}(\beta^{k+1}+\sigma^{-1}w^k).$$
	{\bf Step 3}: Given $\beta^{k+1}$, $u^{k+1}$, $v^{k+1}$, $z^k$ and $w^k$, compute
	$$z^{k+1}=z^k+\tau \sigma (X \beta^{k+1}-u^{k+1}-y), \quad w^{k+1}=w^k+\tau \sigma(\beta^{k+1}-v^{k+1}).$$
\end{framed}

The non-interdependence of the $u$ and $v$ subproblems makes the above iterative framework equivalent to 2-block ADMM with guaranteed convergence. Specific convergence results are detailed in \citet{FPS2013}, and will not be further elaborated here.

\section{Numerical experiments}\label{num}

To thoroughly evaluate the numerical performance of the proposed GSRE method, we compare it with several classic linear regression methods. Specifically, we compare it with well-established ridge regression (Ridge), lasso, adaptive lasso (Alasso) and elastic net (Enet), which do not utilize the structural information of the predictor graph. Their results are obtained by running their corresponding internal functions in MATLAB. Additionally, to highlight the advantages of the graph penalty in this paper, we compare GSRE with the traditional square-root lasso (SRL) in \citet{BCW2011}. Furthermore, to demonstrate the advantages of the square-root loss in this paper, we compare GSRE with the graph sparse regression based on least squares loss, i.e., SRIG in \citet{YL2016}.

\subsection{Simulation data}\label{pc}

In the simulation studies, the predictor graph is defined by the precision matrix of the predictors. Specifically, we employ a MATLAB implementation graphicalLasso.m, which uses a graphical Lasso method \citep{FHT2008} for estimating sparse precision matrix and can be available at \url{https://github.com/xiaohuichen88/Graphical-Lasso/tree/main}. It is worth noting that the precision matrix has already been symmetrized by the previous procedure. In addition, the supplementary materials provide a sensitivity analysis of GSRE to inaccuracies in the estimated graph. We also consider the performance of GSRE and SRIG with the true predictor graph, denoted as GSRE-o and SRIG-o, respectively. In the entire simulation experiment, we uniformly set $p=100$ and the true coefficient vector as $\beta^*=(3,\ldots,3,0,\ldots,0)^{\top}$ and $s^*=15$. For the sensitive parameter $\lambda$, we adopt the grid search method. Solution paths are computed for $\lambda/(\sqrt{n/2}\|X\|)=2^{-13},2^{-12.8},\ldots,2^{-5.2},2^{-5}$, which refers to \citet{BLS2013}. This grid is empirically constructed to cover all potentially interesting solutions. Then we use the HBIC criterion \citep{WKL2013} to select the optimal tuning parameter. The simulated data consists of a training set, an independent validation set, and an independent test set. All models are fitted only on the training data. The validation set is used to choose the appropriate tuning parameter. The test set is used to evaluate the performance of different methods. We use the notation ./././ to denote the sample sizes in the training, validation and test sets, respectively. To highlight $p>n$, for each example, we consider three cases: (I) 20/20/400, (II) 40/40/400 and (III) 60/60/400. For each case, we repeat the simulation 50 times to observe the stability of the algorithms.

To ensure a comprehensive comparison, we consider three different predictor structures and four different types of noise. The methods for generating the predictors are as follows:\\
\textbf{Example 1:} The predictors are generated as:
\begin{align*}
	&X_j=Z_1+0.4\epsilon_j^x, \ Z_1\sim N(0,1), \ 1\le j\le 5;\\
	&X_j=Z_2+0.4\epsilon_j^x, \ Z_2\sim N(0,1), \ 6\le j\le 10;\\
	&X_j=Z_3+0.4\epsilon_j^x, \ Z_3\sim N(0,1), \ 11\le j\le 15;\\
	&X_j\stackrel{i.i.d}{\sim}N(0,1), \ 16\le j \le 100,\\
    &where \ \epsilon_j^x \stackrel{i.i.d}{\sim} N(0,1), \ j=1,2,\ldots,15.
\end{align*}
\textbf{Example 2:} The predictors are generated as:
\begin{align*}
&X_j=Z_1+0.75\epsilon_j^x, \ Z_1\sim Uniform[-1,1], \ 1\le j\le 5;\\
&X_j=Z_2+0.75\epsilon_j^x, \ Z_2\sim Uniform[-1,1], \ 6\le j\le 10;\\
&X_j=Z_3+0.75\epsilon_j^x, \ Z_3\sim Uniform[-1,1], \ 11\le j\le 15;\\
&X_j\stackrel{i.i.d}{\sim}Uniform[-1,1], \ 16\le j \le 100,\\
&where \ \epsilon_j^x \stackrel{i.i.d}{\sim} Uniform[-1,1], \ j=1,2,\ldots,15.
\end{align*}
\textbf{Example 3:} The predictors $(X_1, X_2,\ldots, X_{100})^{\top}\sim N(0,\Sigma)$ with $\Sigma_{ij}=0.5^{|i-j|}$.

\begin{figure}[h]
\begin{center}
\includegraphics[width=5in]{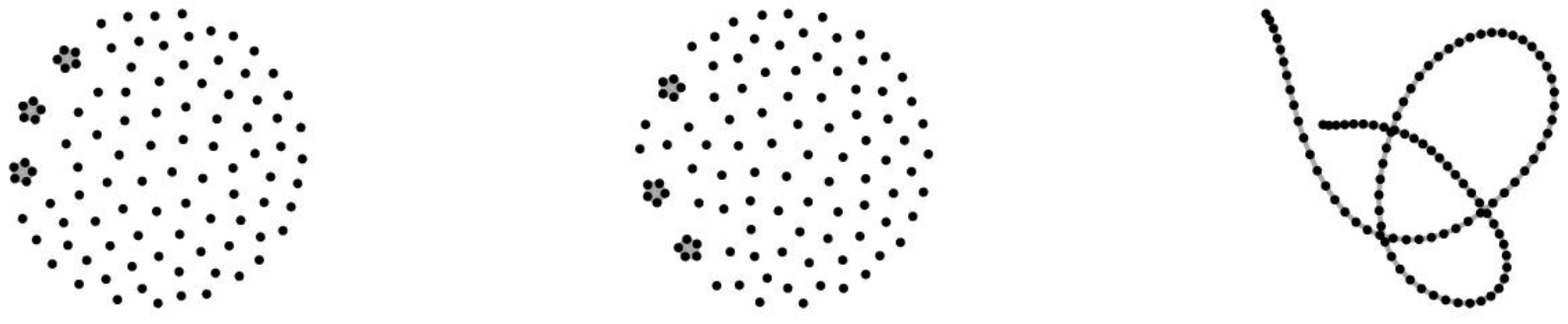}
\end{center}
\caption{True predictor graphs of three simulated examples. \label{fig:graph}}
\end{figure}
Figure \ref{fig:graph} illustrates the true predictor graphs of the three simulated examples. Since Examples 1 and 2 have the same correlation among predictors except for differences in their internal distributions, their predictor graph structures are identical. Since there are three groups of correlations among the first 15 predictors, the predictor graph contains three interconnected subgraphs. Because the remaining predictors are independent of each other, the predictor graph includes 85 isolated nodes. Due to their similarity, we have combined their experimental results. Since the final tables are too large, they are placed at the end of the paper. Assuming the precision matrix corresponding to the predictor graph is denoted by $\Omega=(\omega_{ij})_{i,j=1,2,\ldots,p}$, the generation method for Example 3 leads to $\omega_{ii}=1.333$, $\omega_{ij}=-0.667$ if $|i-j|=1$ and $\omega_{ij}=0$ if $|i-j|>1$. This indicates that predictors are only correlated with variables in their neighborhood, resulting in a banded structure for the predictor graph.

To account for the diversity of noise types, we will describe four different types of noise:
\begin{itemize}
  \item Gaussian distribution: $\epsilon \sim N(0,5^2)$.
  \item T-distribution: $\epsilon \sim t(2)$.
  \item Laplace distribution: $\epsilon \sim Laplace(0,5/\sqrt{2})$.
  \item Uniform distribution: $\epsilon \sim Uniform[-5\sqrt{3},5\sqrt{3}]$.
\end{itemize}
Here, we consider Gaussian distribution with high noise levels, t-distribution with heavier tails, laplace distribution with high kurtosis, and uniform distribution presenting a scenario with constant probability. These all create challenging statistical environments. To evaluate the performances of different methods, we use the following measures:
\begin{itemize}
  \item $L_2$ distance = $\|\hat{\beta}-\beta^*\|_2$;
  \item Relative prediction error (RPE) = $\frac{1}{\sigma^2N_{test}}(\hat{\beta}-\beta^*)^{\top}X_{test}^{\top}X_{test}(\hat{\beta}-\beta^*)$, where $X_{test}$ is the test samples and $N_{test}$ is the number of test samples;
  \item False positive rate (FPR)= $\frac{FP}{FP+TN}$, false negative rate (FNR)= $\frac{FN}{TP+FN}$ and Mattews correlation coefficient (MCC)= $\frac{TP\times TN-FP\times FN}{\sqrt{(TP+FP)(TP+FN)(TN+FP)(TN+FN)}}$, where $TP$, $TN$, $FP$, and $FN$ denote the number of true positives, true negatives, false positives, and false negatives, respectively.
%\begin{align*}
%	&TP=\sum_{j=1}^{p}I(\beta_j^*\ne0)I(\hat{\beta}_j\ne0),\qquad FN=\sum_{j=1}^{p}I(\beta_j^*\ne0)I(\hat{\beta}_j=0),&&\\
%	&TN=\sum_{j=1}^{p}I(\beta_j^*=0)I(\hat{\beta}_j=0),\qquad FP=\sum_{j=1}^{p}I(\beta_j^*=0)I(\hat{\beta}_j\ne0).&&
%\end{align*}
\end{itemize}

Tables \ref{tab:1-1}-\ref{tab:1-4} show the performance of each method for Examples 1 and 2 under four different types of noise. The numbers in parentheses represent the corresponding standard deviations, which are used to measure the stability of these methods. A review of the results in the tables shows that the patterns presented in the four tables are similar. Specifically, among the five methods that do not utilize the predictor graph structure, Enet achieves better estimation results. Alasso performs better in RPE. Although Ridge has the smallest FNR, its FPR is nearly 1. This phenomenon indicates that nearly all zero elements are estimated as non-zero elements. In addition, we find that the results obtained by using the estimated predictor graph are the same as those obtained by using the true predictor graph, which indicates that we have achieved accurate estimation of the graph structure between predictors in these simulation examples. In cases (I) and (II), where $p\gg n$, the proposed GSRE method generally performs best in estimation, prediction and model selection for Example 1 and Example 2. For case (III), where the sample size is slightly larger, GSRE remains optimal in estimation and prediction, but is slightly inferior to SRIG in model selection.

Tables \ref{tab:3-1}-\ref{tab:3-4} display the performance of each method for Example 3 under four different types of noise. Similarly, as the sample size increases, the numerical performance of each algorithm improves. Ridge often gives a low FNR but a high FPR, which indicate a tendency towards overfitting. Lasso and Alasso methods generally have lower values of FPR, but they have higher values for $L_2$ distance and FNR. Enet still performs best in the methods that don't use the graph structure between predictors. In all four challenging environments, the GSRE method consistently outperforms other methods in model estimation and prediction, which shows its robust performance. By balancing the values of FPR, FNR and MCC, it is clear that our method also performs very well in model selection for Example 3.
%%%%%%%%%%%%%%%%%%%%m2n1%%%%%%%%%%%%%%%%%%%%%%%%%%%%%%%%%%%%%%%%%%%%%%%%%%%%%
\begin{table}[H]
	\caption{Performance comparison of estimation, prediction and model selection for Example 3 with Gaussian noise.}
	\centering{\footnotesize
	\label{tab:3-1}
    \vspace{10pt}
	\begin{tabular}{lllllllllll}
		\hline
		& &  & \multicolumn{1}{c}{$L_2$ distance}  & \multicolumn{1}{c}{RPE} & \multicolumn{1}{c}{FPR}  & \multicolumn{1}{c}{FNR} & \multicolumn{1}{c}{MCC}\\
		\hline
		\multirow{7}{*}{\makecell[c]{(I)\\20/20/400}}
        & Ridge   & & 8.181(1.355) & 5.224(2.651) & 0.948(0.022) & 0.010(0.027) & 0.071(0.056) \\
        & Lasso   & & 10.790(1.225) & 6.233(2.605) & 0.072(0.019) & 0.489(0.142) & 0.452(0.142) \\
        & Alasso  & & 10.963(1.113) & 6.352(2.476) & 0.051(0.018) & 0.509(0.148) & 0.484(0.153) \\
        & Enet    & & 8.000(1.268) & 3.221(1.878) & 0.138(0.027) & 0.152(0.108) & 0.591(0.105) \\
        & SRL     & & 11.259(1.259) & 6.894(2.434) & 0.116(0.029) & 0.481(0.124) & 0.379(0.126) \\
        & SRIG-o  & & 10.393(0.359) & 11.403(1.097) & 0.852(0.060) & 0.019(0.049) & 0.139(0.051) \\
        & SRIG    & & 10.393(0.359) & 11.403(1.097) & 0.852(0.060) & 0.019(0.049) & 0.139(0.051) \\
        & GSRE-o  & & 7.095(1.540) & 4.613(2.541) & 0.318(0.105) & 0.031(0.074) & 0.483(0.125) \\
        & GSRE    & & 7.095(1.540) & 4.613(2.541) & 0.318(0.105) & 0.031(0.074) & 0.483(0.125) \\
		\hline
		\multirow{7}{*}{\makecell[c]{(II)\\40/40/400}}
        & Ridge   & & 7.755(0.560) & 3.800(1.064) & 0.954(0.017) & 0.000(0.000) & 0.083(0.016) \\
        & Lasso   & & 8.186(1.035) & 3.724(1.461) & 0.028(0.018) & 0.280(0.107) & 0.732(0.096) \\
        & Alasso  & & 8.039(1.238) & 2.137(0.898) & 0.034(0.024) & 0.225(0.085) & 0.756(0.091) \\
        & Enet    & & 6.827(0.945) & 3.940(1.385) & 0.118(0.037) & 0.059(0.071) & 0.692(0.079) \\
        & SRL     & & 8.236(1.174) & 3.898(1.517) & 0.028(0.018) & 0.289(0.121) & 0.726(0.108) \\
        & SRIG-o  & & 7.649(1.432) & 6.121(2.508) & 0.335(0.200) & 0.001(0.009) & 0.531(0.220) \\
        & SRIG    & & 7.649(1.432) & 6.121(2.508) & 0.335(0.200) & 0.001(0.009) & 0.531(0.220) \\
        & GSRE-o  & & 5.145(1.057) & 2.129(1.316) & 0.039(0.033) & 0.013(0.038) & 0.887(0.079) \\
        & GSRE    & & 5.145(1.057) & 2.129(1.316) & 0.039(0.033) & 0.013(0.038) & 0.887(0.079) \\
		\hline
		\multirow{7}{*}{\makecell[c]{(III)\\60/60/400}}
        & Ridge   & & 6.819(0.620) & 2.354(0.521) & 0.957(0.011) & 0.000(0.000) & 0.081(0.011) \\
        & Lasso   & & 6.615(1.597) & 2.682(1.693) & 0.019(0.017) & 0.180(0.160) & 0.829(0.096) \\
        & Alasso  & & 6.796(1.099) & 1.310(0.557) & 0.026(0.019) & 0.141(0.082) & 0.833(0.082) \\
        & Enet    & & 6.074(1.148) & 3.389(1.448) & 0.068(0.031) & 0.037(0.062) & 0.801(0.065) \\
        & SRL     & & 6.061(1.247) & 2.011(1.082) & 0.029(0.022) & 0.105(0.110) & 0.849(0.087) \\
        & SRIG-o  & & 4.911(0.862) & 2.182(0.918) & 0.025(0.018) & 0.000(0.000) & 0.926(0.045) \\
        & SRIG    & & 4.911(0.862) & 2.182(0.918) & 0.025(0.018) & 0.000(0.000) & 0.926(0.045) \\
        & GSRE-o  & & 4.266(0.914) & 1.454(0.731) & 0.023(0.023) & 0.001(0.009) & 0.934(0.057) \\
        & GSRE    & & 4.266(0.914) & 1.454(0.731) & 0.023(0.023) & 0.001(0.009) & 0.934(0.057) \\
		\hline
	\end{tabular}}
\end{table}
%%%%%%%%%%%%%%%%%%%%%%%%%%%%%%%%%%%m2n2%%%%%%%%%%%%%%%%%%%%%%%%%%%%%%%%%%%%%%%
\begin{table}[H]
	\caption{Performance comparison of estimation, prediction and model selection for Example 3 with T-distribution noise.}
	\centering{\footnotesize
	\label{tab:3-2}
    \vspace{10pt}
	\begin{tabular}{lllllllllll}
		\hline
		& &  & \multicolumn{1}{c}{$L_2$ distance} &  \multicolumn{1}{c}{RPE}&  \multicolumn{1}{c}{FPR} &  \multicolumn{1}{c}{FNR}    & \multicolumn{1}{c}{MCC}         \\
		\hline
		\multirow{7}{*}{\makecell[c]{(I)\\20/20/400}}
        & Ridge    && 7.503(1.441) & 4.224(2.498) & 0.953(0.017) & 0.001(0.009) & 0.082(0.024) \\
        & Lasso    && 9.299(2.075) & 4.521(2.763) & 0.062(0.022) & 0.375(0.196) & 0.564(0.177) \\
        & Alasso   && 10.301(1.703) & 5.345(2.686) & 0.044(0.020) & 0.457(0.176) & 0.547(0.161) \\
        & Enet     && 6.492(2.081) & 2.348(2.108) & 0.116(0.027) & 0.113(0.129) & 0.653(0.129) \\
        & SRL      && 10.548(1.437) & 5.809(2.439) & 0.113(0.021) & 0.413(0.133) & 0.435(0.129) \\
        & SRIG-o   && 10.370(0.355) & 11.352(1.084) & 0.842(0.062) & 0.023(0.050) & 0.141(0.049) \\
        & SRIG     && 10.370(0.355) & 11.352(1.084) & 0.842(0.062) & 0.023(0.050) & 0.141(0.049) \\
        & GSRE-o   && 6.067(2.006) & 3.474(2.631) & 0.308(0.096) & 0.027(0.063) & 0.494(0.122) \\
        & GSRE     && 6.067(2.006) & 3.474(2.631) & 0.308(0.096) & 0.027(0.063) & 0.494(0.122) \\
		\hline
		\multirow{7}{*}{\makecell[c]{(II)\\40/40/400}}
        & Ridge    && 7.048(0.941) & 3.185(1.268) & 0.951(0.013) & 0.000(0.000) & 0.087(0.012) \\
        & Lasso    && 5.718(2.606) & 2.039(2.176) & 0.040(0.035) & 0.140(0.166) & 0.798(0.130) \\
        & Alasso   && 4.648(3.015) & 1.051(2.080) & 0.026(0.037) & 0.081(0.129) & 0.878(0.146) \\
        & Enet     && 5.550(1.828) & 2.768(2.040) & 0.119(0.053) & 0.043(0.073) & 0.705(0.098) \\
        & SRL      && 5.911(2.682) & 2.231(2.324) & 0.040(0.035) & 0.148(0.177) & 0.794(0.144) \\
        & SRIG-o   && 6.748(1.884) & 4.865(2.962) & 0.235(0.216) & 0.000(0.000) & 0.650(0.246) \\
        & SRIG     && 6.748(1.884) & 4.865(2.962) & 0.235(0.216) & 0.000(0.000) & 0.650(0.246) \\
        & GSRE-o   && 3.463(1.554) & 1.067(1.625) & 0.046(0.034) & 0.007(0.034) & 0.872(0.085) \\
        & GSRE     && 3.463(1.554) & 1.067(1.625) & 0.046(0.034) & 0.007(0.034) & 0.872(0.085) \\
		\hline
		\multirow{7}{*}{\makecell[c]{(III)\\60/60/400}}
        & Ridge    && 5.677(0.966) & 1.624(0.804) & 0.958(0.012) & 0.000(0.000) & 0.080(0.012) \\
        & Lasso    && 3.497(1.866) & 0.744(1.205) & 0.043(0.029) & 0.025(0.093) & 0.866(0.106) \\
        & Alasso   && 3.378(4.223) & 0.834(3.931) & 0.018(0.049) & 0.027(0.087) & 0.938(0.135) \\
        & Enet     && 3.995(1.654) & 1.479(1.530) & 0.108(0.044) & 0.015(0.049) & 0.742(0.086) \\
        & SRL      && 3.528(2.249) & 0.790(1.540) & 0.047(0.048) & 0.028(0.093) & 0.861(0.134) \\
        & SRIG-o   && 4.421(1.037) & 1.801(1.175) & 0.032(0.059) & 0.000(0.000) & 0.919(0.081) \\
        & SRIG     && 4.421(1.037) & 1.801(1.175) & 0.032(0.059) & 0.000(0.000) & 0.919(0.081) \\
        & GSRE-o   && 2.720(1.205) & 0.618(0.960) & 0.033(0.035) & 0.001(0.009) & 0.911(0.081) \\
        & GSRE     && 2.720(1.205) & 0.618(0.960) & 0.033(0.035) & 0.001(0.009) & 0.911(0.081) \\
		\hline
	\end{tabular}}
\end{table}

%%%%%%%%%%%%%%%%%%%%%%%%%%%%%%%%%%%%%%m2n3%%%%%%%%%%%%%%%%%%%%%%%%%%%%%%%%%%%%%
\begin{table}[H]
	\caption{Performance comparison of estimation, prediction and model selection for Example 3 with Laplace noise.}
	\centering{\footnotesize
	\label{tab:3-3}
    \vspace{10pt}
	\begin{tabular}{llllllllllll}
		\hline
		& &  & \multicolumn{1}{c}{$L_2$ distance}  & \multicolumn{1}{c}{RPE} & \multicolumn{1}{c}{FPR}  & \multicolumn{1}{c}{FNR}    & \multicolumn{1}{c}{MCC}              \\
		\hline
		\multirow{7}{*}{\makecell[c]{(I)\\20/20/400}}
        & Ridge    && 8.072(1.420) & 5.017(2.618) & 0.953(0.018) & 0.011(0.025) & 0.065(0.050) \\
        & Lasso    && 10.352(1.464) & 5.791(2.887) & 0.071(0.021) & 0.456(0.177) & 0.480(0.178) \\
        & Alasso   && 11.112(1.386) & 6.363(2.460) & 0.050(0.022) & 0.531(0.141) & 0.471(0.168) \\
        & Enet     && 7.837(1.450) & 3.158(1.804) & 0.133(0.027) & 0.155(0.113) & 0.596(0.113) \\
        & SRL      && 10.967(1.373) & 6.487(2.826) & 0.118(0.033) & 0.475(0.134) & 0.382(0.145) \\
        & SRIG-o   && 10.390(0.337) & 11.386(1.029) & 0.839(0.060) & 0.021(0.049) & 0.144(0.055) \\
        & SRIG     && 10.390(0.337) & 11.386(1.029) & 0.839(0.060) & 0.021(0.049) & 0.144(0.055) \\
        & GSRE-o   && 6.915(1.644) & 4.334(2.494) & 0.309(0.105) & 0.027(0.069) & 0.493(0.102) \\
        & GSRE     && 6.915(1.644) & 4.334(2.494) & 0.309(0.105) & 0.027(0.069) & 0.493(0.102) \\
		\hline
		\multirow{7}{*}{\makecell[c]{(II)\\40/40/400}}
        & Ridge    && 7.653(0.645) & 3.759(1.161) & 0.952(0.016) & 0.000(0.000) & 0.085(0.015) \\
        & Lasso    && 7.975(1.209) & 3.643(1.610) & 0.028(0.018) & 0.265(0.124) & 0.743(0.104) \\
        & Alasso   && 7.553(1.284) & 1.896(0.861) & 0.029(0.020) & 0.204(0.111) & 0.782(0.113) \\
        & Enet     && 6.653(0.994) & 3.791(1.404) & 0.115(0.029) & 0.060(0.066) & 0.693(0.078) \\
        & SRL      && 8.001(1.266) & 3.875(1.831) & 0.026(0.020) & 0.284(0.143) & 0.737(0.110) \\
        & SRIG-o   && 7.675(1.516) & 6.179(2.520) & 0.338(0.197) & 0.003(0.019) & 0.522(0.211) \\
        & SRIG     && 7.675(1.516) & 6.179(2.520) & 0.338(0.197) & 0.003(0.019) & 0.522(0.211) \\
        & GSRE-o   && 4.919(1.207) & 2.006(1.384) & 0.038(0.035) & 0.005(0.023) & 0.895(0.081) \\
        & GSRE     && 4.919(1.207) & 2.006(1.384) & 0.038(0.035) & 0.005(0.023) & 0.895(0.081) \\
		\hline
		\multirow{7}{*}{\makecell[c]{(III)\\60/60/400}}
        & Ridge    && 6.677(0.751) & 2.267(0.614) & 0.955(0.015) & 0.000(0.000) & 0.083(0.015) \\
        & Lasso    && 6.310(1.666) & 2.446(1.717) & 0.020(0.016) & 0.160(0.149) & 0.838(0.078) \\
        & Alasso   && 6.047(1.358) & 1.053(0.505) & 0.024(0.019) & 0.105(0.076) & 0.862(0.080) \\
        & Enet     && 6.131(1.201) & 3.518(1.546) & 0.057(0.026) & 0.043(0.060) & 0.822(0.051) \\
        & SRL      && 5.790(1.234) & 1.804(0.917) & 0.028(0.017) & 0.085(0.101) & 0.864(0.075) \\
        & SRIG-o   && 4.854(0.701) & 2.101(0.734) & 0.020(0.013) & 0.000(0.000) & 0.939(0.036) \\
        & SRIG     && 4.854(0.701) & 2.101(0.734) & 0.020(0.013) & 0.000(0.000) & 0.939(0.036) \\
        & GSRE-o   && 3.947(0.799) & 1.213(0.627) & 0.021(0.019) & 0.001(0.009) & 0.939(0.050) \\
        & GSRE     && 3.947(0.799) & 1.213(0.627) & 0.021(0.019) & 0.001(0.009) & 0.939(0.050) \\
		\hline
	\end{tabular}}
\end{table}

%%%%%%%%%%%%%%%%%%%%%%%%%%%%%%%%%m2n4%%%%%%%%%%%%%%%%%%%%%%%%%%%%%%%%%%%%%%%%%%
\begin{table}[H]
	\caption{Performance comparison of estimation, prediction and model selection for Example 3 with uniform noise.}
	\centering{\footnotesize
	\label{tab:3-4}
    \vspace{10pt}
	\begin{tabular}{llllllllll}
		\hline
		& &  & \multicolumn{1}{c}{$L_2$ distance}  & \multicolumn{1}{c}{RPE} & \multicolumn{1}{c}{FPR}  & \multicolumn{1}{c}{FNR}  & \multicolumn{1}{c}{MCC}             \\
		\hline
		\multirow{7}{*}{\makecell[c]{(I)\\20/20/400}}
        & Ridge    && 8.127(1.315) & 4.979(2.447) & 0.952(0.019) & 0.007(0.024) & 0.074(0.035) \\
        & Lasso    && 10.704(1.523) & 6.192(2.880) & 0.075(0.021) & 0.487(0.171) & 0.445(0.173) \\
        & Alasso   && 11.246(1.370) & 6.429(2.626) & 0.054(0.020) & 0.527(0.141) & 0.465(0.147) \\
        & Enet     && 8.245(1.245) & 3.518(2.011) & 0.143(0.029) & 0.185(0.117) & 0.560(0.113) \\
        & SRL      && 11.241(1.350) & 6.724(2.731) & 0.117(0.030) & 0.484(0.124) & 0.375(0.130) \\
        & SRIG-o   && 10.393(0.337) & 11.387(1.015) & 0.842(0.065) & 0.024(0.052) & 0.138(0.059) \\
        & SRIG     && 10.393(0.337) & 11.387(1.015) & 0.842(0.065) & 0.024(0.052) & 0.138(0.059) \\
        & GSRE-o   && 7.045(1.609) & 4.483(2.545) & 0.312(0.119) & 0.039(0.080) & 0.484(0.120) \\
        & GSRE     && 7.045(1.609) & 4.483(2.545) & 0.312(0.119) & 0.039(0.080) & 0.484(0.120) \\
		\hline
		\multirow{7}{*}{\makecell[c]{(II)\\40/40/400}}
        & Ridge    && 7.721(0.639) & 3.827(1.099) & 0.955(0.013) & 0.000(0.000) & 0.083(0.013) \\
        & Lasso    && 8.156(1.058) & 3.784(1.538) & 0.029(0.018) & 0.285(0.121) & 0.725(0.108) \\
        & Alasso   && 7.710(1.191) & 2.017(0.872) & 0.034(0.020) & 0.204(0.101) & 0.767(0.102) \\
        & Enet     && 6.757(0.973) & 3.916(1.457) & 0.117(0.025) & 0.064(0.070) & 0.686(0.075) \\
        & SRL      && 8.210(1.197) & 4.045(1.749) & 0.026(0.019) & 0.303(0.147) & 0.722(0.131) \\
        & SRIG-o   && 7.713(1.534) & 6.256(2.657) & 0.342(0.203) & 0.000(0.000) & 0.525(0.219) \\
        & SRIG     && 7.713(1.534) & 6.256(2.657) & 0.342(0.203) & 0.000(0.000) & 0.525(0.219) \\
        & GSRE-o   && 5.059(1.195) & 2.119(1.400) & 0.039(0.035) & 0.011(0.034) & 0.890(0.077) \\
        & GSRE     && 5.059(1.195) & 2.119(1.400) & 0.039(0.035) & 0.011(0.034) & 0.890(0.077) \\
		\hline
		\multirow{7}{*}{\makecell[c]{(III)\\60/60/400}}
        & Ridge    && 6.685(0.683) & 2.366(0.633) & 0.958(0.014) & 0.000(0.000) & 0.080(0.013) \\
        & Lasso    && 6.577(1.508) & 2.670(1.777) & 0.017(0.016) & 0.165(0.156) & 0.845(0.090) \\
        & Alasso   && 6.645(1.177) & 1.256(0.507) & 0.024(0.016) & 0.139(0.084) & 0.842(0.068) \\
        & Enet     && 6.353(1.085) & 3.776(1.478) & 0.053(0.026) & 0.045(0.064) & 0.829(0.057) \\
        & SRL      && 6.123(1.181) & 2.054(0.972) & 0.023(0.015) & 0.103(0.090) & 0.866(0.058) \\
        & SRIG-o   && 4.874(0.794) & 2.123(0.844) & 0.021(0.010) & 0.001(0.009) & 0.936(0.026) \\
        & SRIG     && 4.874(0.794) & 2.123(0.844) & 0.021(0.010) & 0.001(0.009) & 0.936(0.026) \\
        & GSRE-o   && 4.170(0.907) & 1.381(0.769) & 0.017(0.017) & 0.001(0.009) & 0.948(0.045) \\
        & GSRE     && 4.170(0.907) & 1.381(0.769) & 0.017(0.017) & 0.001(0.009) & 0.948(0.045) \\
		\hline
	\end{tabular}}
\end{table}

Besides these three examples, we also provide a complex simulation example and conduct a sensitivity analysis of Assumption \ref{ass2} in the supplementary materials. The numerical results indicate that our proposed GSRE method still outperforms the other methods. In summary, the simulation results indicate that the proposed GSRE method, which combines square-root estimation with graph penalty, shows superior performance in estimation, prediction and model selection under various noise environments. This result highlights the general applicability and reliability of the GSRE method in statistical modeling.

\begin{sidewaystable}[h]
\setlength{\tabcolsep}{1pt}
	\centering{\footnotesize
	\caption{Performance comparison of estimation, prediction and model selection for Example 1\&2 with Gaussian noise.}
	\label{tab:1-1}
 \vspace{10pt}
	\begin{tabular}{lllllllllllll}
		\hline
		&  &\multicolumn{5}{c}{\textbf{Example 1}} &  &\multicolumn{5}{c}{\textbf{Example 2}}\\
		\cline{3-7}\cline{9-13}  \\
		& & \multicolumn{1}{c}{$L_2$ distance} & \multicolumn{1}{c}{RPE}& \multicolumn{1}{c}{FPR} & \multicolumn{1}{c}{FNR}&\multicolumn{1}{c}{MCC}& &\multicolumn{1}{c}{$L_2$ distance}& \multicolumn{1}{c}{RPE}& \multicolumn{1}{c}{FPR} & \multicolumn{1}{c}{FNR}&\multicolumn{1}{c}{MCC}\\
		\hline
		\multirow{7}{*}{\makecell[c]{(I)\\20/20/400}}
    	& Ridge    & 6.833(1.252) & 3.050(3.099) & 0.948(0.020) & 0.001(0.009) & 0.087(0.020) & &  6.820(1.356) & 3.424(2.814) & 0.951(0.016) & 0.000(0.000) & 0.086(0.015) \\
    & Lasso    & 12.841(1.684) & 3.426(2.338) & 0.061(0.024) & 0.508(0.105) & 0.465(0.128) & &  11.335(1.217) & 5.704(3.023) & 0.071(0.023) & 0.497(0.117) & 0.451(0.131) \\
    & Alasso   & 13.044(2.072) & 3.334(2.795) & 0.048(0.024) & 0.497(0.117) & 0.510(0.144) & &  11.449(1.490) & 5.155(2.982) & 0.048(0.021) & 0.476(0.134) & 0.523(0.143) \\
    & Enet     & 9.317(1.443) & 2.473(2.346) & 0.150(0.027) & 0.265(0.105) & 0.491(0.101) & &  8.581(1.239) & 3.164(2.006) & 0.143(0.033) & 0.217(0.095) & 0.538(0.110) \\
    & SRL      & 12.550(1.702) & 3.838(2.701) & 0.123(0.023) & 0.447(0.109) & 0.394(0.106) & &  11.571(1.403) & 5.708(2.832) & 0.130(0.021) & 0.440(0.114) & 0.388(0.117) \\
    & SRIG-o   & 8.819(0.661) & 13.752(2.380) & 0.576(0.061) & 0.000(0.000) & 0.317(0.036) & &  9.235(0.516) & 12.075(1.678) & 0.562(0.058) & 0.011(0.049) & 0.316(0.046) \\
    & SRIG     & 8.819(0.661) & 13.752(2.380) & 0.576(0.061) & 0.000(0.000) & 0.317(0.036) & &  9.233(0.519) & 12.071(1.682) & 0.562(0.058) & 0.011(0.049) & 0.316(0.046) \\
    & GSRE-o   & 3.219(1.057) & 1.392(1.772) & 0.128(0.048) & 0.000(0.000) & 0.718(0.074) & &  4.319(1.165) & 1.941(2.111) & 0.160(0.051) & 0.007(0.047) & 0.666(0.070) \\
    & GSRE     & 3.219(1.057) & 1.392(1.772) & 0.128(0.048) & 0.000(0.000) & 0.718(0.074) & &  4.320(1.164) & 1.941(2.111) & 0.160(0.051) & 0.007(0.047) & 0.666(0.071) \\
		\hline
		\multirow{7}{*}{\makecell[c]{(II)\\40/40/400}}
	& Ridge    & 6.325(0.695) & 2.078(0.929) & 0.966(0.018) & 0.000(0.000) & 0.071(0.019) & &  6.813(0.879) & 2.864(1.108) & 0.962(0.015) & 0.000(0.000) & 0.076(0.015) \\
    & Lasso    & 9.765(1.470) & 2.401(1.123) & 0.019(0.015) & 0.353(0.110) & 0.709(0.088) & &  8.063(1.345) & 2.946(1.329) & 0.022(0.017) & 0.255(0.130) & 0.768(0.105) \\
    & Alasso   & 9.758(1.469) & 0.984(0.440) & 0.024(0.017) & 0.268(0.099) & 0.751(0.096) & &  8.184(1.392) & 1.603(0.750) & 0.025(0.016) & 0.213(0.105) & 0.787(0.093) \\
    & Enet     & 4.696(0.852) & 2.998(1.419) & 0.105(0.031) & 0.000(0.000) & 0.753(0.053) & &  5.668(0.913) & 3.400(1.454) & 0.118(0.030) & 0.019(0.049) & 0.718(0.066) \\
    & SRL      & 9.554(1.531) & 2.459(1.392) & 0.016(0.014) & 0.333(0.116) & 0.732(0.097) & &  7.970(1.372) & 3.042(1.553) & 0.024(0.015) & 0.249(0.114) & 0.766(0.090) \\
    & SRIG-o   & 3.300(1.290) & 2.000(1.967) & 0.038(0.067) & 0.000(0.000) & 0.914(0.124) & &  3.944(1.283) & 2.144(1.927) & 0.045(0.067) & 0.000(0.000) & 0.896(0.127) \\
    & SRIG     & 3.300(1.290) & 2.000(1.967) & 0.038(0.067) & 0.000(0.000) & 0.914(0.124) & &  3.945(1.282) & 2.142(1.928) & 0.045(0.067) & 0.000(0.000) & 0.896(0.127) \\
    & GSRE-o   & 2.431(0.616) & 0.751(0.598) & 0.009(0.010) & 0.000(0.000) & 0.972(0.029) & &  3.011(0.613) & 0.811(0.524) & 0.014(0.011) & 0.000(0.000) & 0.956(0.031) \\
    & GSRE     & 2.431(0.616) & 0.751(0.598) & 0.009(0.010) & 0.000(0.000) & 0.972(0.029) & &  3.014(0.613) & 0.811(0.524) & 0.014(0.011) & 0.000(0.000) & 0.956(0.031) \\
		\hline
		\multirow{7}{*}{\makecell[c]{(III)\\60/60/400}}
    & Ridge    & 6.267(0.781) & 2.272(0.605) & 0.957(0.014) & 0.000(0.000) & 0.081(0.014) & &  6.348(0.777) & 2.285(0.596) & 0.956(0.013) & 0.000(0.000) & 0.082(0.012) \\
    & Lasso    & 8.110(1.492) & 1.497(0.941) & 0.010(0.010) & 0.223(0.121) & 0.829(0.081) & &  6.211(1.549) & 1.725(1.214) & 0.009(0.011) & 0.108(0.135) & 0.906(0.082) \\
    & Alasso   & 9.619(1.312) & 0.736(0.230) & 0.020(0.012) & 0.261(0.072) & 0.772(0.061) & &  7.092(1.272) & 1.042(0.418) & 0.018(0.014) & 0.135(0.079) & 0.862(0.069) \\
    & Enet     & 3.903(0.613) & 2.137(0.856) & 0.055(0.030) & 0.000(0.000) & 0.855(0.069) & &  4.715(1.023) & 2.579(1.403) & 0.047(0.029) & 0.011(0.037) & 0.865(0.076) \\
    & SRL      & 7.791(1.332) & 1.248(0.615) & 0.013(0.012) & 0.172(0.105) & 0.851(0.075) & &  5.931(1.453) & 1.489(0.998) & 0.013(0.013) & 0.072(0.099) & 0.916(0.074) \\
    & SRIG-o   & 2.269(0.446) & 0.734(0.385) & 0.004(0.007) & 0.000(0.000) & 0.987(0.022) & &  2.701(0.583) & 0.792(0.461) & 0.001(0.004) & 0.000(0.000) & 0.996(0.011) \\
    & SRIG     & 2.269(0.446) & 0.734(0.385) & 0.004(0.007) & 0.000(0.000) & 0.987(0.022) & &  2.706(0.582) & 0.792(0.461) & 0.001(0.004) & 0.000(0.000) & 0.996(0.011) \\
    & GSRE-o   & 2.028(0.341) & 0.413(0.241) & 0.005(0.007) & 0.000(0.000) & 0.984(0.022) & &  2.464(0.477) & 0.511(0.306) & 0.003(0.006) & 0.000(0.000) & 0.991(0.019) \\
    & GSRE     & 2.028(0.341) & 0.413(0.241) & 0.005(0.007) & 0.000(0.000) & 0.984(0.022) & &  2.468(0.476) & 0.511(0.306) & 0.003(0.006) & 0.000(0.000) & 0.991(0.019) \\
		\hline
	\end{tabular}}
\end{sidewaystable}

\begin{sidewaystable}[h]
\setlength{\tabcolsep}{1pt}
	\caption{Performance comparison of estimation, prediction and model selection for Example 1\&2 with T-distribution noise.}
	\centering{\footnotesize
	\label{tab:1-2}
\vspace{10pt}
	\begin{tabular}{lllllllllllll}
		\hline
		&  &\multicolumn{5}{c}{\textbf{Example 1}} &  &\multicolumn{5}{c}{\textbf{Example 2}}\\
		\cline{3-7}\cline{9-13}  \\
		& & \multicolumn{1}{c}{$L_2$ distance} & \multicolumn{1}{c}{RPE}& \multicolumn{1}{c}{FPR} & \multicolumn{1}{c}{FNR}&\multicolumn{1}{c}{MCC}& &\multicolumn{1}{c}{$L_2$ distance}& \multicolumn{1}{c}{RPE}& \multicolumn{1}{c}{FPR} & \multicolumn{1}{c}{FNR}&\multicolumn{1}{c}{MCC}\\
		\hline
		\multirow{7}{*}{\makecell[c]{(I)\\20/20/400}}
& Ridge    & 6.148(1.203) & 2.124(2.710) & 0.948(0.019) & 0.000(0.000) & 0.089(0.017) & &  6.104(1.347) & 2.683(2.486) & 0.953(0.016) & 0.001(0.009) & 0.082(0.025)\\
& Lasso    & 10.737(1.884) & 2.528(2.470) & 0.039(0.022) & 0.405(0.144) & 0.606(0.153) & &  9.997(1.916) & 4.221(2.683) & 0.058(0.027) & 0.408(0.146) & 0.553(0.160) \\
& Alasso   & 10.612(2.038) & 1.937(2.471) & 0.026(0.024) & 0.373(0.142) & 0.669(0.160) & &  9.908(2.205) & 4.109(2.916) & 0.036(0.023) & 0.403(0.165) & 0.614(0.162) \\
& Enet     & 7.129(2.365) & 1.344(1.761) & 0.116(0.036) & 0.192(0.136) & 0.599(0.144) & &  6.830(1.950) & 1.922(1.587) & 0.113(0.030) & 0.159(0.111) & 0.627(0.118) \\
& SRL      & 10.439(1.944) & 2.440(2.333) & 0.108(0.024) & 0.337(0.147) & 0.504(0.136) & &  10.056(1.668) & 4.310(2.286) & 0.113(0.026) & 0.344(0.136) & 0.490(0.142) \\
& SRIG-o   & 8.822(0.654) & 13.761(2.319) & 0.559(0.062) & 0.000(0.000) & 0.326(0.037) & &  9.213(0.500) & 12.015(1.639) & 0.550(0.054) & 0.005(0.030) & 0.328(0.034) \\
& SRIG     & 8.822(0.654) & 13.761(2.319) & 0.559(0.062) & 0.000(0.000) & 0.326(0.037) & &  9.211(0.504) & 12.010(1.645) & 0.550(0.054) & 0.005(0.030) & 0.328(0.034) \\
& GSRE-o   & 2.280(0.762) & 0.450(0.835) & 0.124(0.037) & 0.000(0.000) & 0.722(0.059) & &  3.127(1.211) & 0.936(1.616) & 0.156(0.053) & 0.005(0.038) & 0.672(0.075) \\
& GSRE     & 2.280(0.762) & 0.450(0.835) & 0.124(0.037) & 0.000(0.000) & 0.722(0.059) & &  3.127(1.211) & 0.936(1.616) & 0.156(0.053) & 0.005(0.038) & 0.672(0.075) \\
		\hline
		\multirow{7}{*}{\makecell[c]{(II)\\40/40/400}}
& Ridge    & 4.991(1.161) & 1.443(1.676) & 0.962(0.019) & 0.000(0.000) & 0.075(0.020) & &   5.440(0.887) & 1.856(0.849) & 0.969(0.016) & 0.000(0.000) & 0.067(0.019) \\
& Lasso    & 6.917(2.493) & 1.288(1.422) & 0.020(0.022) & 0.168(0.167) & 0.833(0.126) & &  5.360(2.606) & 1.553(1.825) & 0.024(0.019) & 0.119(0.151) & 0.854(0.101)\\
& Alasso   & 6.379(2.773) & 0.591(1.225) & 0.011(0.027) & 0.121(0.142) & 0.894(0.149) & &  4.424(2.505) & 0.577(0.669) & 0.010(0.012) & 0.068(0.107) & 0.925(0.081) \\
& Enet     & 3.667(1.299) & 1.713(1.899) & 0.110(0.054) & 0.001(0.009) & 0.750(0.088) & &  4.323(1.399) & 1.930(1.813) & 0.119(0.053) & 0.006(0.019) & 0.732(0.084) \\
& SRL      & 6.838(2.383) & 1.159(1.202) & 0.021(0.022) & 0.148(0.140) & 0.843(0.121) & &  5.249(2.256) & 1.261(1.102) & 0.028(0.020) & 0.097(0.121) & 0.857(0.083) \\
& SRIG-o   & 2.661(1.171) & 1.345(2.103) & 0.008(0.039) & 0.000(0.000) & 0.983(0.074) & &  3.022(0.827) & 1.163(0.882) & 0.005(0.021) & 0.000(0.000) & 0.987(0.049) \\
& SRIG     & 2.661(1.171) & 1.345(2.103) & 0.008(0.039) & 0.000(0.000) & 0.983(0.074) & &  3.020(0.829) & 1.160(0.884) & 0.005(0.021) & 0.000(0.000) & 0.987(0.049) \\
& GSRE-o   & 1.749(0.874) & 0.382(0.997) & 0.009(0.009) & 0.000(0.000) & 0.972(0.028) & &  1.872(0.723) & 0.280(0.479) & 0.015(0.011) & 0.000(0.000) & 0.954(0.032) \\
& GSRE     & 1.749(0.874) & 0.382(0.997) & 0.009(0.009) & 0.000(0.000) & 0.972(0.028) & &  1.872(0.723) & 0.280(0.479) & 0.015(0.011) & 0.000(0.000) & 0.954(0.032) \\
		\hline
		\multirow{7}{*}{\makecell[c]{(III)\\60/60/400}}
   & Ridge    & 4.661(0.562) & 1.193(0.437) & 0.954(0.017) & 0.000(0.000) & 0.083(0.018) & &  5.243(0.981) & 1.642(0.973) & 0.954(0.016) & 0.000(0.000) & 0.084(0.016) \\
& Lasso    & 4.041(1.313) & 0.279(0.190) & 0.016(0.015) & 0.015(0.037) & 0.944(0.048) & &  3.524(2.231) & 0.758(1.529) & 0.018(0.021) & 0.043(0.130) & 0.920(0.112) \\
& Alasso   & 3.966(2.020) & 0.139(0.129) & 0.003(0.006) & 0.045(0.080) & 0.964(0.051) & &  3.485(3.752) & 0.658(2.436) & 0.016(0.064) & 0.031(0.084) & 0.950(0.156) \\
& Enet     & 2.294(0.455) & 0.476(0.247) & 0.081(0.040) & 0.000(0.000) & 0.802(0.078) & &  3.170(1.341) & 1.105(1.343) & 0.077(0.054) & 0.007(0.034) & 0.809(0.106) \\
& SRL      & 3.801(1.291) & 0.251(0.184) & 0.022(0.017) & 0.011(0.031) & 0.929(0.051) & &  3.335(2.029) & 0.552(0.904) & 0.029(0.061) & 0.024(0.085) & 0.913(0.150) \\
& SRIG-o    & 1.815(0.333) & 0.483(0.197) & 0.001(0.004) & 0.000(0.000) & 0.996(0.011) & &  2.247(0.744) & 0.614(0.587) & 0.005(0.023) & 0.000(0.000) & 0.989(0.054) \\
& SRIG     & 1.815(0.333) & 0.483(0.197) & 0.001(0.004) & 0.000(0.000) & 0.996(0.011) & &  2.250(0.742) & 0.615(0.587) & 0.005(0.023) & 0.000(0.000) & 0.989(0.054) \\
& GSRE-o   & 1.331(0.297) & 0.091(0.070) & 0.006(0.010) & 0.000(0.000) & 0.980(0.030) & &  1.591(0.688) & 0.194(0.330) & 0.008(0.021) & 0.000(0.000) & 0.979(0.050) \\
& GSRE     & 1.331(0.297) & 0.091(0.070) & 0.006(0.010) & 0.000(0.000) & 0.980(0.030) & &  1.591(0.688) & 0.194(0.330) & 0.008(0.021) & 0.000(0.000) & 0.979(0.050) \\
		\hline
	\end{tabular}}
\end{sidewaystable}

%%%%%%%%%%%%%%%%%%%%%%%%%%%%m1n3-m4n3%%%%%%%%%%%%%%%%%%%%%%%%%%%%%%%%%%%%%%%%%%%
\begin{sidewaystable}[h]
\setlength{\tabcolsep}{1pt}
	\caption{Performance comparison of estimation, prediction and model selection for Example 1\&2 with Laplace noise.}
	\centering{\footnotesize
	\label{tab:1-3}
\vspace{10pt}
	\begin{tabular}{lllllllllllll}
		\hline
		&  &\multicolumn{5}{c}{\textbf{Example 1}} &  &\multicolumn{5}{c}{\textbf{Example 2}}\\
		\cline{3-7}\cline{9-13}  \\
		& & \multicolumn{1}{c}{$L_2$ distance} & \multicolumn{1}{c}{RPE}& \multicolumn{1}{c}{FPR} & \multicolumn{1}{c}{FNR}&\multicolumn{1}{c}{MCC}& &\multicolumn{1}{c}{$L_2$ distance}& \multicolumn{1}{c}{RPE}& \multicolumn{1}{c}{FPR} & \multicolumn{1}{c}{FNR}&\multicolumn{1}{c}{MCC}\\
		\hline
		\multirow{7}{*}{\makecell[c]{(I)\\20/20/400}}
  & Ridge    & 6.584(1.155) & 2.749(2.811) & 0.947(0.021) & 0.000(0.000) & 0.089(0.019) & &  6.826(1.336) & 3.229(2.910) & 0.952(0.017) & 0.001(0.009) & 0.083(0.022) \\
& Lasso    & 12.095(1.424) & 3.178(2.209) & 0.059(0.028) & 0.487(0.097) & 0.491(0.133) & &  11.222(1.616) & 4.903(3.013) & 0.071(0.019) & 0.445(0.123) & 0.491(0.125) \\
& Alasso   & 12.167(1.531) & 2.970(3.385) & 0.037(0.023) & 0.464(0.109) & 0.567(0.135) & &  10.992(1.709) & 4.591(2.836) & 0.051(0.020) & 0.431(0.137) & 0.551(0.140) \\
& Enet     & 9.386(1.700) & 2.662(2.712) & 0.148(0.033) & 0.285(0.116) & 0.481(0.118) & &  8.474(1.316) & 2.774(1.608) & 0.140(0.030) & 0.211(0.106) & 0.546(0.108) \\
& SRL      & 12.388(1.566) & 3.751(3.120) & 0.125(0.024) & 0.451(0.116) & 0.387(0.115) & &  11.617(1.812) & 5.419(2.589) & 0.128(0.022) & 0.429(0.125) & 0.399(0.126) \\
& SRIG-o   & 8.867(0.652) & 13.907(2.361) & 0.568(0.069) & 0.007(0.047) & 0.316(0.048) & &  9.196(0.560) & 11.974(1.763) & 0.568(0.058) & 0.004(0.021) & 0.318(0.033) \\
& SRIG     & 8.867(0.652) & 13.907(2.361) & 0.568(0.069) & 0.007(0.047) & 0.316(0.048) & &  9.194(0.565) & 11.968(1.772) & 0.568(0.058) & 0.004(0.021) & 0.318(0.033) \\
& GSRE-o   & 3.022(0.890) & 1.145(1.178) & 0.129(0.035) & 0.000(0.000) & 0.713(0.058) & &  4.029(1.139) & 1.518(1.721) & 0.153(0.047) & 0.007(0.047) & 0.675(0.069) \\
& GSRE     & 3.022(0.890) & 1.145(1.178) & 0.129(0.035) & 0.000(0.000) & 0.713(0.058) & &  4.029(1.139) & 1.518(1.721) & 0.153(0.047) & 0.007(0.047) & 0.675(0.069) \\
		\hline
		\multirow{7}{*}{\makecell[c]{(II)\\40/40/400}}
& Ridge    & 6.059(0.749) & 2.098(1.317) & 0.965(0.021) & 0.000(0.000) & 0.072(0.020) & &  6.953(0.855) & 2.900(1.155) & 0.964(0.016) & 0.001(0.009) & 0.071(0.023) \\
& Lasso    & 9.571(1.366) & 2.585(1.339) & 0.014(0.013) & 0.333(0.104) & 0.741(0.098) & &  8.235(1.267) & 3.047(1.656) & 0.023(0.018) & 0.261(0.116) & 0.761(0.102) \\
& Alasso   & 9.571(1.582) & 0.988(0.541) & 0.016(0.016) & 0.239(0.100) & 0.799(0.091) & &  8.483(1.328) & 1.763(0.912) & 0.029(0.019) & 0.209(0.097) & 0.779(0.096) \\
& Enet     & 4.760(0.970) & 3.227(1.599) & 0.100(0.035) & 0.003(0.013) & 0.761(0.063) & &  5.722(1.018) & 3.432(1.689) & 0.126(0.034) & 0.019(0.036) & 0.705(0.065) \\
& SRL      & 9.297(1.569) & 2.495(1.577) & 0.018(0.019) & 0.304(0.131) & 0.748(0.111) & &  8.059(1.234) & 3.106(1.921) & 0.023(0.018) & 0.253(0.128) & 0.765(0.112) \\
& SRIG-o   & 3.140(1.194) & 1.783(1.760) & 0.024(0.052) & 0.000(0.000) & 0.943(0.099) & &  4.341(1.509) & 2.696(2.218) & 0.076(0.088) & 0.000(0.000) & 0.839(0.161) \\
& SRIG     & 3.140(1.194) & 1.783(1.760) & 0.024(0.052) & 0.000(0.000) & 0.943(0.099) & &  4.338(1.503) & 2.687(2.201) & 0.076(0.088) & 0.000(0.000) & 0.840(0.162) \\
& GSRE-o   & 2.380(0.575) & 0.702(0.593) & 0.011(0.011) & 0.000(0.000) & 0.966(0.033) & &  3.046(0.727) & 0.834(0.634) & 0.014(0.016) & 0.000(0.000) & 0.957(0.044) \\
& GSRE     & 2.380(0.575) & 0.702(0.593) & 0.011(0.011) & 0.000(0.000) & 0.966(0.033) & &  3.051(0.730) & 0.833(0.633) & 0.014(0.016) & 0.000(0.000) & 0.957(0.044) \\
		\hline
		\multirow{7}{*}{\makecell[c]{(III)\\60/60/400}}
& Ridge    & 6.078(0.724) & 2.109(0.569) & 0.958(0.014) & 0.000(0.000) & 0.080(0.015) & &  6.470(0.743) & 2.411(0.783) & 0.958(0.012) & 0.000(0.000) & 0.080(0.012) \\
& Lasso    & 7.711(1.308) & 1.454(0.721) & 0.007(0.011) & 0.196(0.118) & 0.858(0.078) & &  6.464(1.403) & 1.901(1.465) & 0.010(0.011) & 0.129(0.127) & 0.889(0.077) \\
& Alasso   & 8.876(1.556) & 0.662(0.279) & 0.013(0.015) & 0.224(0.089) & 0.818(0.089) & &  7.267(1.357) & 1.110(0.504) & 0.021(0.017) & 0.153(0.082) & 0.840(0.081) \\
& Enet     & 3.858(0.585) & 2.108(0.805) & 0.052(0.025) & 0.000(0.000) & 0.861(0.058) & &  4.840(1.090) & 2.678(1.641) & 0.047(0.022) & 0.008(0.029) & 0.866(0.055) \\
& SRL      & 7.476(1.269) & 1.212(0.600) & 0.012(0.012) & 0.153(0.098) & 0.866(0.074) & &  6.220(1.298) & 1.642(0.958) & 0.015(0.013) & 0.093(0.085) & 0.895(0.067) \\
& SRIG-o   & 2.295(0.414) & 0.767(0.334) & 0.002(0.005) & 0.000(0.000) & 0.994(0.016) & &  2.789(0.562) & 0.829(0.474) & 0.003(0.006) & 0.000(0.000) & 0.991(0.019) \\
& SRIG     & 2.295(0.414) & 0.767(0.334) & 0.002(0.005) & 0.000(0.000) & 0.994(0.016) & &  2.794(0.553) & 0.829(0.474) & 0.003(0.006) & 0.000(0.000) & 0.991(0.019) \\
& GSRE-o   & 2.055(0.331) & 0.453(0.226) & 0.004(0.007) & 0.000(0.000) & 0.988(0.021) & &  2.543(0.424) & 0.518(0.269) & 0.006(0.009) & 0.000(0.000) & 0.980(0.028) \\
& GSRE     & 2.055(0.331) & 0.453(0.226) & 0.004(0.007) & 0.000(0.000) & 0.988(0.021) & &  2.547(0.418) & 0.518(0.269) & 0.006(0.009) & 0.000(0.000) & 0.980(0.028) \\
		\hline
	\end{tabular}}
\end{sidewaystable}

%%%%%%%%%%%%%%%%%%%%%%%%%%%%%%%%%%%%%%%%%%%%%%%%%%%%m1n4-m4n4%%%%%%%%%%%%%%%%%%%%%%%%%%%%%%%%%%%%
\begin{sidewaystable}[h]
\setlength{\tabcolsep}{1pt}
	\caption{Performance comparison of estimation, prediction and model selection for Example 1\&2 with uniform noise.}
	\centering{\footnotesize
	\label{tab:1-4}
\vspace{10pt}
	\begin{tabular}{lllllllllllll}
		\hline
		&  &\multicolumn{5}{c}{\textbf{Example 1}} &  &\multicolumn{5}{c}{\textbf{Example 2}}\\
		\cline{3-7}\cline{9-13}  \\
		& & \multicolumn{1}{c}{$L_2$ distance} & \multicolumn{1}{c}{RPE}& \multicolumn{1}{c}{FPR} & \multicolumn{1}{c}{FNR}&\multicolumn{1}{c}{MCC}& &\multicolumn{1}{c}{$L_2$ distance}& \multicolumn{1}{c}{RPE}& \multicolumn{1}{c}{FPR} & \multicolumn{1}{c}{FNR}&\multicolumn{1}{c}{MCC}\\
		\hline
		\multirow{7}{*}{\makecell[c]{(I)\\20/20/400}}
& Ridge    & 6.636(1.193) & 2.672(2.646) & 0.952(0.016) & 0.003(0.013) & 0.081(0.029) & &  6.724(1.175) & 2.989(2.248) & 0.955(0.014) & 0.001(0.009) & 0.080(0.028) \\
& Lasso    & 12.319(1.584) & 3.090(2.225) & 0.064(0.032) & 0.477(0.110) & 0.490(0.141) & &  11.318(1.622) & 5.450(3.316) & 0.067(0.025) & 0.467(0.133) & 0.485(0.148) \\
& Alasso   & 12.449(1.728) & 3.101(3.412) & 0.044(0.024) & 0.476(0.110) & 0.537(0.136) & &  11.236(1.514) & 4.815(2.851) & 0.053(0.021) & 0.448(0.136) & 0.533(0.147) \\
& Enet     & 9.433(1.293) & 2.506(2.153) & 0.151(0.027) & 0.284(0.102) & 0.477(0.103) & &  8.590(1.360) & 2.847(2.307) & 0.139(0.027) & 0.216(0.096) & 0.544(0.100) \\
& SRL      & 12.583(1.426) & 3.788(3.163) & 0.128(0.020) & 0.452(0.123) & 0.381(0.119) & &  11.673(1.773) & 5.453(2.772) & 0.128(0.028) & 0.435(0.138) & 0.395(0.140) \\
& SRIG-o   & 8.866(0.659) & 13.900(2.364) & 0.572(0.071) & 0.007(0.047) & 0.314(0.050) & &  9.216(0.540) & 12.024(1.726) & 0.558(0.061) & 0.004(0.021) & 0.324(0.034) \\
& SRIG     & 8.866(0.659) & 13.900(2.364) & 0.572(0.071) & 0.007(0.047) & 0.314(0.050) & &  9.215(0.542) & 12.022(1.730) & 0.558(0.061) & 0.004(0.021) & 0.324(0.034) \\
& GSRE-o   & 3.198(1.048) & 1.371(1.624) & 0.138(0.042) & 0.000(0.000) & 0.701(0.065) & &  4.123(1.163) & 1.564(1.840) & 0.144(0.040) & 0.007(0.047) & 0.687(0.062) \\
& GSRE     & 3.198(1.048) & 1.371(1.624) & 0.138(0.042) & 0.000(0.000) & 0.701(0.065) & &  4.124(1.162) & 1.564(1.840) & 0.144(0.040) & 0.007(0.047) & 0.687(0.062) \\
		\hline
		\multirow{7}{*}{\makecell[c]{(II)\\40/40/400}}
& Ridge    & 6.257(0.727) & 2.131(1.253) & 0.965(0.016) & 0.000(0.000) & 0.072(0.018) & &  6.886(0.836) & 2.852(1.260) & 0.964(0.017) & 0.001(0.009) & 0.069(0.030) \\
& Lasso    & 9.781(1.313) & 2.646(1.412) & 0.016(0.015) & 0.352(0.105) & 0.720(0.097) & &  8.167(1.232) & 2.913(1.622) & 0.024(0.015) & 0.239(0.108) & 0.773(0.096) \\
& Alasso   & 9.858(1.525) & 1.022(0.461) & 0.021(0.013) & 0.253(0.103) & 0.770(0.091) & &  8.351(1.316) & 1.634(0.793) & 0.028(0.014) & 0.205(0.088) & 0.784(0.076) \\
& Enet     & 4.743(0.911) & 3.097(1.494) & 0.109(0.036) & 0.003(0.013) & 0.746(0.063) & &  5.658(0.952) & 3.311(1.576) & 0.127(0.031) & 0.017(0.040) & 0.704(0.059) \\
& SRL      & 9.483(1.613) & 2.447(1.423) & 0.020(0.017) & 0.337(0.127) & 0.717(0.104) & &  8.017(1.287) & 3.007(1.883) & 0.023(0.015) & 0.237(0.119) & 0.775(0.092) \\
& SRIG-o   & 3.270(1.305) & 1.989(2.091) & 0.034(0.065) & 0.000(0.000) & 0.924(0.120) & &  4.196(1.332) & 2.438(1.887) & 0.070(0.089) & 0.000(0.000) & 0.854(0.160) \\
& SRIG     & 3.270(1.305) & 1.989(2.091) & 0.034(0.065) & 0.000(0.000) & 0.924(0.120) & &  4.192(1.330) & 2.426(1.889) & 0.069(0.090) & 0.000(0.000) & 0.856(0.161) \\
& GSRE-o   & 2.446(0.584) & 0.736(0.602) & 0.012(0.012) & 0.000(0.000) & 0.965(0.036) & &  2.945(0.579) & 0.724(0.481) & 0.016(0.015) & 0.000(0.000) & 0.953(0.042) \\
& GSRE     & 2.446(0.584) & 0.736(0.602) & 0.012(0.012) & 0.000(0.000) & 0.965(0.036) & &  2.950(0.584) & 0.724(0.480) & 0.016(0.015) & 0.000(0.000) & 0.953(0.042) \\
		\hline
		\multirow{7}{*}{\makecell[c]{(III)\\60/60/400}}
& Ridge    & 6.174(0.742) & 2.215(0.538) & 0.957(0.016) & 0.000(0.000) & 0.080(0.015) & &  6.426(0.720) & 2.424(0.832) & 0.959(0.011) & 0.000(0.000) & 0.079(0.012) \\
& Lasso    & 8.016(1.343) & 1.559(0.765) & 0.008(0.010) & 0.229(0.114) & 0.830(0.074) & &  6.463(1.353) & 1.917(1.353) & 0.010(0.013) & 0.140(0.126) & 0.883(0.077) \\
& Alasso   & 9.167(1.393) & 0.698(0.263) & 0.011(0.012) & 0.243(0.079) & 0.814(0.078) & &  7.247(1.284) & 1.080(0.461) & 0.017(0.014) & 0.157(0.085) & 0.848(0.078) \\
& Enet     & 3.956(0.664) & 2.250(0.949) & 0.052(0.025) & 0.000(0.000) & 0.859(0.057) & &  4.776(1.123) & 2.602(1.686) & 0.049(0.027) & 0.009(0.036) & 0.862(0.068) \\
& SRL      & 7.647(1.211) & 1.252(0.551) & 0.013(0.013) & 0.164(0.098) & 0.858(0.065) & &  6.158(1.142) & 1.598(0.878) & 0.016(0.016) & 0.089(0.089) & 0.897(0.060) \\
& SRIG-o   & 2.341(0.423) & 0.789(0.357) & 0.002(0.004) & 0.000(0.000) & 0.994(0.014) & &  2.796(0.590) & 0.846(0.516) & 0.002(0.005) & 0.000(0.000) & 0.993(0.016) \\
& SRIG     & 2.341(0.423) & 0.789(0.357) & 0.002(0.004) & 0.000(0.000) & 0.994(0.014) & &  2.802(0.584) & 0.846(0.516) & 0.002(0.005) & 0.000(0.000) & 0.993(0.016) \\
& GSRE-o   & 2.112(0.340) & 0.463(0.229) & 0.005(0.008) & 0.000(0.000) & 0.985(0.023) & &  2.516(0.453) & 0.507(0.312) & 0.006(0.010) & 0.000(0.000) & 0.983(0.030) \\
& GSRE     & 2.112(0.340) & 0.463(0.229) & 0.005(0.008) & 0.000(0.000) & 0.985(0.023) & &  2.521(0.450) & 0.508(0.312) & 0.006(0.010) & 0.000(0.000) & 0.983(0.030) \\
		\hline
	\end{tabular}}
\end{sidewaystable}

\subsection{Real data}
In this part, we further evaluate the numerical performance of the GSRE method by two real data sets.

The first set is the bodyfat2 data, which originates from the bodyfat data in the LIBSVM database. This data is suitable for large-scale regression problems and can be available at \url{https://www.csie.ntu.edu.tw/~cjlin/libsvmtools/datasets}. For this study, we first preprocess the data by removing zero columns. Additionally, we use the method of \citet{HJY2010} to expand the data by applying polynomial basis functions to the original features. The final digit 2 in bodyfat2 indicates that we use second-order polynomials to generate the basis functions. After expansion, we obtain the bodyfat2 data with a sample size of 252 and a dimensionality of 120. To satisfy the high-dimensional assumption, i.e., $p>n$, we randomly select 100 samples. Therefore, the size of the data set we employ is $100\times120$.

The second set is the miRNA data, which is a small noncoding RNAs that play a crucial role in various biological processes such as development, cellular differentiation, cell cycle regulation and programmed cell death. Recent studies have shown a strong association between miRNA and cancer \citep{HJP2014,KWF2012,GRC2009}. Here, we use the miRNA expression data for thymic cancer, which can be downloaded from R package.
This data is sourced from thymoma tissue samples in the Cancer Genome Atlas database. This data set provides miRNA expression values for 113 patients across 1,881 miRNAs. After removing duplicate features and filtering out low-expression genes, we obtain expression values for 136 miRNAs. Consequently, the data size is $113\times136$, which satisfies the high-dimensional assumption of $p > n$. We also consider survival time as the response variable $y$.

\begin{figure}[h]
\centering
\subfloat[bodyfat2]{%
\includegraphics[width=0.5\textwidth]{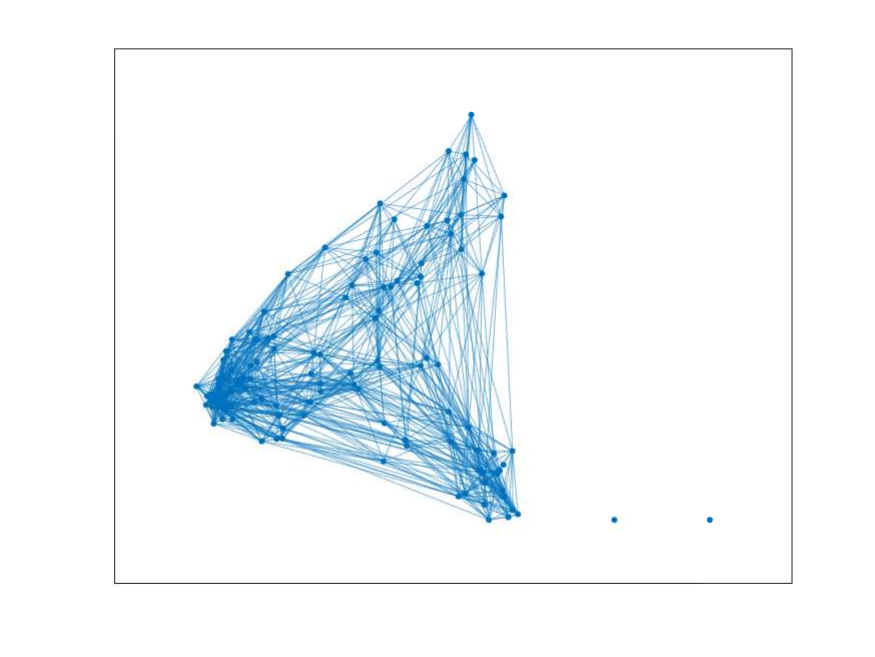}%
\label{fig:global_data}%
}\hfill
\subfloat[miRNA]{%
\includegraphics[width=0.5\textwidth]{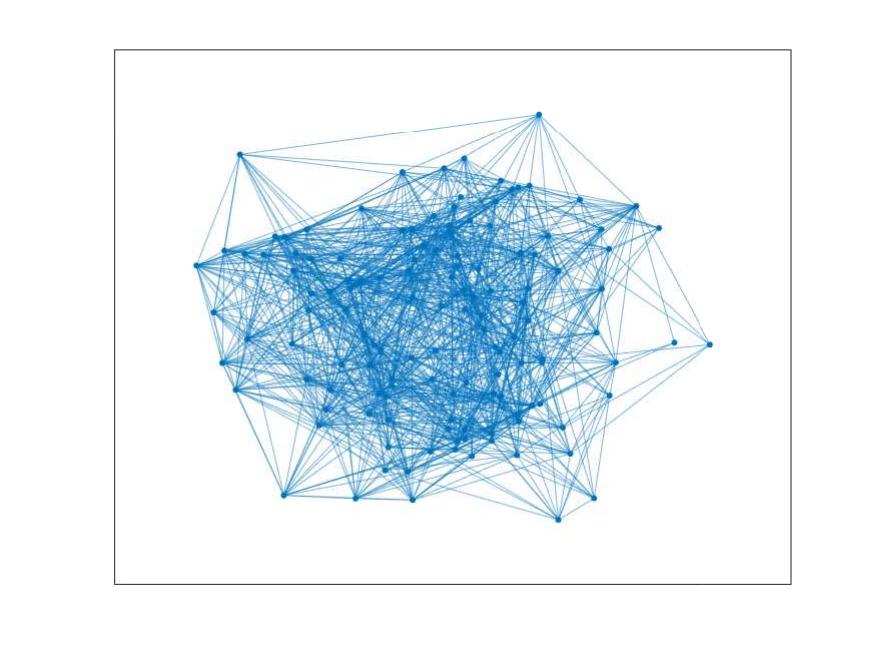}%
\label{fig:long_term_data}%
}
\caption{\itshape Estimated graphs of real data.}
\label{fig:11}
\end{figure}
Before inputting the preprocessed data into the algorithm for iteration, we perform normalization on the data sets. We then use the graphical lasso \citep{FHT2008} to estimate the graphical structure of the predictors, as detailed in Figure \ref{fig:11}. It is evident that there is a high degree of correlation among the predictors in both data sets.

\begin{figure}[h]
\centering
\subfloat[bodyfat2]{%
\includegraphics[width=0.5\textwidth]{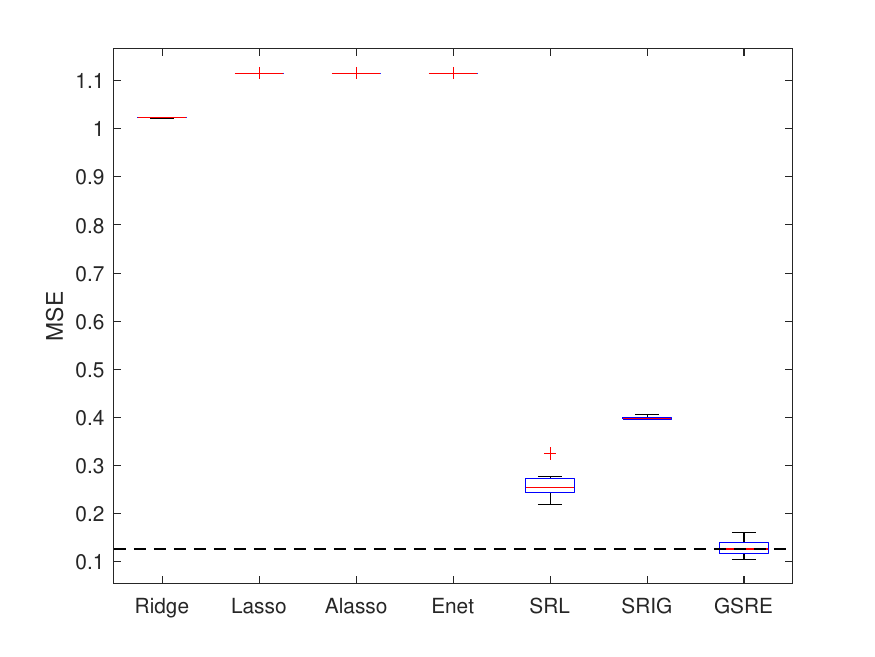}%
\label{fig:global_data}%
}\hfill
\subfloat[miRNA]{%
\includegraphics[width=0.5\textwidth]{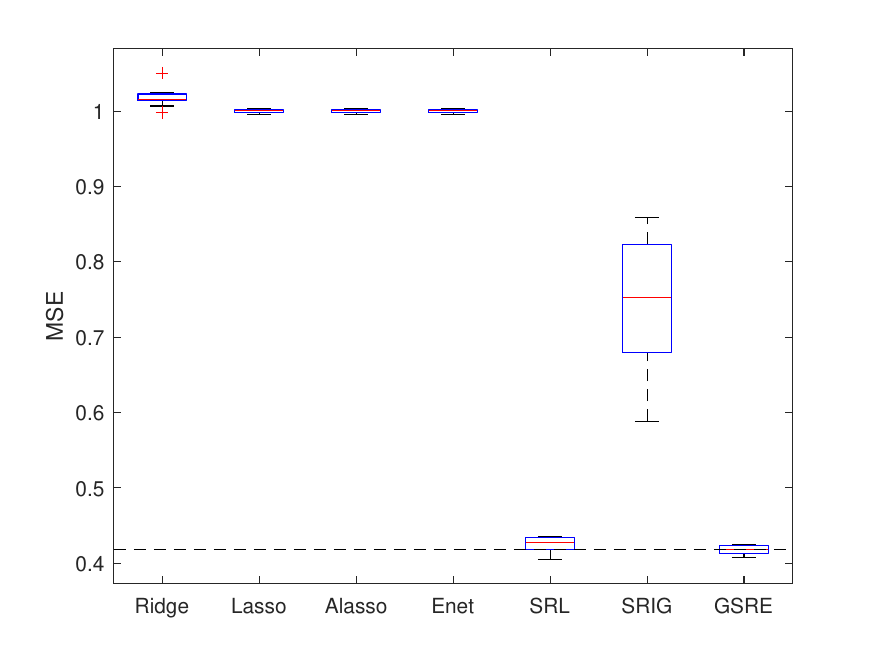}%
\label{fig:long_term_data}%
}
\caption{\itshape Comparison of MSE for various methods on real data.}
\label{fig:22}
\end{figure}

In the specific experiments, we use ten-fold cross-validation to evaluate different methods. Here, we randomly select nine folds of data as the training set and use the remaining data as the test set. All models are fitted on the training set and evaluated on the test data. To select the tuning parameters for different methods, we use five-fold cross-validation on the training set. The experiment was repeated 10 times to avoid biases caused by random partitioning. Here, we uniformly use mean squared error (MSE) to measure prediction accuracy. Figure \ref{fig:22} shows box plots of MSE for different methods on two real data sets.
%with the calculation format as follows:
%$$MSE:=\frac{1}{n_{test}}\sum_{i=1}^{n_{test}}(y_i^{test}-\langle X_i^{test},\hat{\beta}\rangle)^2,$$
%where $y_i^{test}$ and $X_i^{test}, i=1,2,\ldots,n_{test}$ are from the test set.

Figure \ref{fig:22}(a) illustrates the performance of various methods on the bodyfa2 data. The results indicate that GSRE significantly outperforms the other six methods in terms of MSE scores. Figure \ref{fig:22}(b) presents the numerical performance of different methods on the miRNA data. This figure clearly shows that the performance of GSRE is significantly superior to that of Ridge, Lasso, Alasso, Enet and SRIG. Although the boxes of GSRE and SRL are both at the bottom, the median of GSRE is lower than that of SRL. These results indicate that the GSRE method has superior numerical performance compared to other methods and is an ideal choice for achieving accurate predictions.

\section{Conclusion}\label{con}
%%%%%%%%%%%%%%%%%%%%%%%%%%%%%%%%%%%%%%%%%%%%%%%%%%%%%%%%%%%
In this paper, we addresses the high-dimensional sparse linear regression problem by proposing a graph-based square-root estimation (GSRE) method, which integrates square-root loss and node-wise graph penalty. This paper highlights the broad applicability of the proposed method by analyzing its equivalence with several classical square-root estimation models. A notable feature of the theoretical analysis is that the tuning parameter no longer depends on the unknown standard deviation of the error terms. Additionally, we establish theoretical results on finite sample bounds and model selection consistency. In terms of computation, we use the fast and efficient alternating direction method of multipliers. Extensive simulation and real tests demonstrate that the proposed GSRE method is a competitive tool for estimation, prediction and model selection. Future work will consider extending this approach to more general loss functions, such as $\ell_p$ ($p\geq1$) loss \citep{W2013,XKLQ2018,ZW2015,WCN2018}, which will facilitate the handling of a broader range of noise types in high-dimensional settings.

\section*{Disclosure statement}

The authors report there are no competing interests to declare.

\section*{Acknowledgments}

The authors thank the Editor, Associate Editor and two anonymous referees for their
helpful comments and suggestions. The authors contributed equally to this paper and are listed in alphabetical order.

\section*{Supplementary materials}

\begin{description}
\item[Appendix:] Contains complete proofs of the theoretical results. Contains detailed explanations of some contents in the paper. Contains some additional simulation examples. (appendix.pdf, a pdf file)
\item[Code:] Contains code that implements the proposed method and reproduces the numerical results. A README file describing the contents is included. (code.zip, a zip file)
\end{description}

\bibliographystyle{Chicago}
\bibliography{references}
\end{document}